\pdfoutput=1

\documentclass[twocolumn,tighten]{aastex62}
\usepackage{amssymb,amsmath,mathrsfs}
\usepackage{ucs}
\usepackage[utf8x]{inputenc}
\usepackage[encapsulated]{CJK}

\newcommand{\cntext}[1]{\begin{CJK}{UTF8}{bkai}#1\ignorespacesafterend\end{CJK}}
\newcommand{\pc}{\textsc{Pencil Code}}

\renewcommand{\vec}[1]{\boldsymbol{#1}}
\newcommand{\pder}[2]{\frac{\partial #1}{\partial #2}}
\newcommand{\pdder}[2]{\frac{\partial^2 #1}{\partial #2^2}}
\newcommand{\tder}[2]{\frac{\mathrm{d}#1}{\mathrm{d}#2}}

\newcommand{\unitvec}{\hat{\vec{e}}}
\newcommand{\del}{\vec{\nabla}}
\newcommand{\dotp}{\boldsymbol{\cdot}}
\newcommand{\crossp}{\boldsymbol{\times}}
\newcommand{\divgc}[1]{\del\dotp#1}
\newcommand{\curl}[1]{\del\crossp#1}
\newcommand{\laplacian}[1]{\nabla^2 #1}
\newcommand{\ha}[2]{\langle #1\rangle_#2}

\newcommand{\Bext}{\vec{B}_\mathrm{ext}}
\newcommand{\OmegaK}{\Omega_\mathrm{K}}
\newcommand{\ReM}{\mathrm{Re}_\mathrm{M}}
\newcommand{\ass}{\alpha_\mathrm{SS}}
\newcommand{\ssa}{Shakura--Sunyaev}

\received{April~10, 2018}
\revised{September~26, 2018}
\accepted{October~9, 2018}
\submitjournal{The Astrophysical Journal}

\shorttitle{Diffusion and Concentration of Solids in Dead Zone}
\shortauthors{Yang, Mac Low, \& Johansen}

\begin{document}
\title{Diffusion and Concentration of Solids in the Dead Zone of a Protoplanetary Disk}

\author[0000-0003-2589-5034]{Chao-Chin Yang (\cntext{楊朝欽})}
\affiliation{%
    Lund Observatory,
    Department of Astronomy and Theoretical Physics, Lund University,
    Box~43, SE-221\,00~Lund, Sweden}
\affiliation{%
    Department of Physics and Astronomy,
    University of Nevada, Las Vegas,
    4505 S.~Maryland Pkwy,
    Box~454002,
    Las Vegas, NV~89154-4002, U.S.A.}

\author[0000-0003-0064-4060]{Mordecai-Mark Mac Low}
\affiliation{%
    Department of Astrophysics,
    American Museum of Natural History,
    Central Park West at 79th Street,
    New York, NY~10024-5192, U.S.A.}
\affiliation{%
    Center for Computational Astrophysics, Flatiron Institute,
    New York, NY, U.S.A.}

\author[0000-0002-5893-6165]{Anders Johansen}
\affiliation{%
    Lund Observatory,
    Department of Astronomy and Theoretical Physics,
    Lund University,
    Box~43, SE-221\,00~Lund, Sweden}

\correspondingauthor{Chao-Chin Yang}
\email{ccyang@unlv.edu}

\begin{abstract}
The streaming instability is a promising mechanism to drive the formation of planetesimals in protoplanetary disks.
To trigger this process, it has been argued that sedimentation of solids onto the mid-plane needs to be efficient and therefore that a quiescent gaseous environment is required.
It is often suggested that dead-zone or disk-wind structure created by non-ideal magnetohydrodynamical (MHD) effects meets this requirement.
However, simulations have shown that the midplane of a dead zone is not completely quiescent.
In order to examine the concentration of solids in such an environment, we use the local-shearing-box approximation to simulate a particle-gas system with an Ohmic dead zone including mutual drag force between the gas and the solids.
We systematically compare the evolution of the system with ideal or non-ideal MHD, with or without back-reaction drag force from particles on gas, and with varying solid abundances.
Similar to previous investigations of dead zone dynamics, we find that particles of dimensionless stopping time $\tau_s=0.1$ do not sediment appreciably more than those in ideal magneto-rotational turbulence, resulting in a vertical scale height an order of magnitude larger than in a laminar disk.
Contrary to the expectation that this should curb the formation of planetesimals, we nevertheless find that strong clumping of solids still occurs in the dead zone when solid abundances are similar to the critical value for a laminar environment.
This can be explained by the weak radial diffusion of particles near the mid-plane.
The results imply that the sedimentation of particles to the mid-plane is not a necessary criterion for the formation of planetesimals by the streaming instability.
\end{abstract}

\keywords{
    instabilities ---
    magnetohydrodynamics (MHD) ---
    methods:~numerical ---
    planets and satellites:~formation ---
    protoplanetary disks ---
    turbulence}

\section{INTRODUCTION}

Planet formation occurs in gaseous protoplanetary disks containing solid materials around young stars.
The process must proceed from interstellar \micron-sized dust grains all the way up to planetary cores, which covers a range of 13 orders of magnitude in size, or almost 40 orders of magnitude in mass.
It also needs to be efficient so that gas giant planets can form before the gaseous disk disperses within about 1--10\,Myr \citep[see, e.g.,][]{WC11}.
In the process, dust particles as well as the ensuing progressively larger bodies in the protoplanetary disk intimately interact with the gas via drag and gravitational forces.
Therefore, their ability to consolidate and form planets is inevitably dictated by the dynamics of the surrounding gas.

It is believed that protoplanetary disks must be at least weakly magnetized, and the very existence of the magnetic fields drives complicated gas dynamics and produces a rich structure within these disks \citep[see, e.g.,][and references therein]{TF14}.
In the inner region ($\lesssim1$\,au) of the disk, the ionization degree is high due to its high temperature, and the magnetically-coupled, differentially-rotating gas is subject to the magneto-rotational instability \citep[MRI;][]{BH91}.
This instability drives turbulence that produces magnetic energy from orbital shear, allowing disk accretion by magnetic stresses.
Further outwards in the disk, the ionization degree in the mid-plane is so low that the MRI becomes inactive, leading to a quasi-quiescent region called a dead zone that may be sandwiched by MRI-active, turbulent surface layers \citep{cG96,FS03}.
If one considers additional non-ideal magnetohydrodynamical (MHD) effects, i.e., ambipolar diffusion and Hall drift, rather than active layers, a magneto-centrifugal wind is launched near the surface of the protoplanetary disk that is dominant in driving disk accretion \citep{xB14,LKF14,GT15}.
In any case, the viscosity near the mid-plane can be two orders of magnitude lower than that in fully developed magneto-rotational turbulence.

The low macroscopic viscosity near the mid-plane of the protoplanetary disk is often argued to imply that the environment for planet formation is effectively laminar.
However, numerical simulations of non-ideal MHD disks indicate that appreciable kinetic energy remains present in the gas near the mid-plane \citep{FP06,OH11,SB13,xB15,GT15}.
In general, the density and velocity fluctuations in the gas near the mid-plane can be $\sim$1--3\% of the mid-plane density and of the local speed of sound, respectively; these fluctuations are believed to be driven by waves propagating down from the turbulent surface layers into the mid-plane \citep{OM09,xB15}.
Even though the fluctuations are weaker than those in fully developed magneto-rotational turbulence, they may still substantially exceed the magnitude that the measured viscous stress would suggest.
The distinction occurs because these motions do not have the correlations expected for magneto-rotational turbulent flow.
Instead, the gas motions in non-ideal MHD can be fairly different in vertical and horizontal directions \citep{ZSB15}.

These fluctuations in the gas near the mid-plane of the disk directly affect the dynamics of the embedded solid bodies. 
The density fluctuations can drive random walks in the orbital properties---including semimajor axis, eccentricity, and inclination---of kilometer-scale planetesimals or larger objects via stochastic gravitational force \citep{YMM09,YMM12,NG10,GNT11,OO13}.
The velocity fluctuations can drive significant random velocities in mm--cm-sized pebbles via frictional drag force \citep{FP06,JO07,BT09,OH11,ZSB15,XBM17,RL18}.
Therefore, from the point of view of solid objects in protoplanetary disks with non-ideal MHD effects, the gas flow surrounding them should still be considered significantly fluctuating, even if not classically turbulent.

It remains unclear how kilometer-scale planetesimals are formed in such an environment inside a dead zone.
One promising mechanism to drive the formation of planetesimals from mm--cm-sized pebbles is the streaming instability, with which these solid particles assist in concentrating themselves via the back reaction to the gas drag \citep{YG05,YJ07,JY07}.
Without externally driven velocity fluctuations, it has been shown that the combination of particle sedimentation and the streaming instability in the nonlinear stage can concentrate solid particles to high densities, as long as enough solids are present in the local column \citep{JYM09,BS10,YJ14,CJD15,YJC17}.
On the other hand, studies of the streaming instability in externally driven fluctuating flows have been sparse.
\cite{JO07,JKH11} showed that distributed particles with dimensionless stopping times $\tau_s = 0.1$--1 (dm--m-sized boulders in the terrestrial region of a solar nebula; e.g., \citealt{JB14}) in ideal magneto-rotational turbulence can concentrate themselves to high densities.
\cite{BT09} and \cite{TB10} did not see strong clumping of solids with a range of particle sizes from \micron{} to cm in a similar environment.
So far, no study of the streaming instability incorporating non-ideal MHD driving of the flow has yet been conducted.

Therefore, we consider in this work the streaming instability, i.e., a particle-gas system with mutual drag interaction, inside an (Ohmic) dead zone of a protoplanetary disk.
We systematically compare the behavior between ideal and non-ideal MHD, with and without back reaction to the gas drag, and with varying solid abundances.
In Section~\ref{S:method}, we describe in detail our models and numerical methods.
We analyze the vertical profiles of the gas properties in the saturated state of ideal and non-ideal MHD flows and measure the diffusion coefficients of the gas in Section~\ref{S:ssgas}.
We study in Section~\ref{S:sspar} the vertical distribution and radial diffusion of the solid particles when no back reaction is in effect and compare the results with analytical expectations.
In Section~\ref{S:csbr}, we activate the back reaction and systematically increase the solid abundance until we find strong concentration of solid materials.
We conclude in Section~\ref{S:cr} with discussion of the implications of this work.

\section{METHODOLOGY} \label{S:method}

\subsection{Governing Equations}

To model a magnetized, gaseous protoplanetary disk loaded with solid materials, we adopt the standard local-shearing-box approximation \citep{GL65,BN95,HGB95}.
This approximation assumes that the dimensions of the computational domain are much smaller than its distance to the central star.\footnote{We note that this approximation is in favor of locations \emph{closer} to the central star, when the ratio of gas scale height $H_g$ to radial distance $R$ increases with increasing $R$.}
The system can then be linearized such that the domain becomes rectilinear with its center revolving around the central star at its local Keplerian angular frequency $\OmegaK$ and with its three axes pointing along the radial, azimuthal, and vertical directions, respectively.
Using this approximation, we describe our governing equations for the MHD and the solid particles in the following subsections.

\subsubsection{Magnetohydrodynamics} \label{SSS:MHD}

We consider gas dynamics in the Eulerian frame.
The gas density $\rho_g$ and velocity $\vec{u}$ are defined on a fixed, regular grid, where $\vec{u}$ is measured relative to the background shear flow $-3\OmegaK x\unitvec_y / 2$.
For simplicity, we adopt the isothermal equation of state with the speed of sound being $c_s$.
To account for the radial pressure gradient in the disk on larger scales, we impose a constant, background radial acceleration $2\Pi c_s\OmegaK\unitvec_x$ on the gas.
The dimensionless coefficient $\Pi \equiv \Delta u_y / c_s$ was defined by \citet{BS10}, with $\Delta u_y$ being the (positive) reduction of the azimuthal gas velocity due the radial pressure gradient.
We also include a constant, uniform, external, vertical magnetic field $\Bext = B_\mathrm{ext}\unitvec_z$.
The continuity and the momentum equations for the gas then read
\begin{align}
&\pder{\rho_g}{t}
    - \frac{3}{2}\OmegaK x\pder{\rho_g}{y}
    + \del\cdot(\rho_g\vec{u})
    = 0,\label{E:mhd_cont}\\
&\pder{\vec{u}}{t}
    - \frac{3}{2}\OmegaK x\pder{\vec{u}}{y}
    + \vec{u}\cdot\del\vec{u}
    = 2\Pi c_s\OmegaK\unitvec_x
    - c_s^2\del\ln\rho_g\nonumber\\
&\qquad+ \left(2\OmegaK u_y\unitvec_x
        - \frac{1}{2}\OmegaK u_x\unitvec_y
        - \OmegaK^2 z\unitvec_z\right)\nonumber\\
&\qquad+ \frac{1}{\rho_g}\vec{J}\times
        \left(\vec{B} + \Bext\right)
    + \frac{\rho_p}{\rho_g}
      \frac{\vec{\tilde{v}} - \vec{u}}{t_s}.\label{E:mhd_mom}
\end{align}
The terms in the parentheses on the right-hand side of Equation~\eqref{E:mhd_mom} are the combination of the linearized gravity from the central star (both horizontal and vertical), the centrifugal force, and the Coriolis force.
The following term is the Lorentz force, where $\vec{B}$ is the magnetic field in addition to $\vec{B}_\mathrm{ext}$, $\vec{J} = \curl{\vec{B}} / \mu_0$ is the current density, and $\mu_0$ is the permeability of the vacuum.
The last term is the back reaction of the drag force exerted on the gas from the solid particles, where $\rho_p$ and $\tilde{\vec{v}}$ is the average density and velocity of the particles contributed to the respective cell of gas \citep{YJ07,YJ16}, and $t_s$ is the stopping time of the drag force (\citealt{fW72,sW77a}; see also Section~\ref{SSS:par}).
In terms of the magnetic vector potential $\vec{A}$, which is also defined on the grid, the induction equation we consider is
\begin{equation}
\pder{\vec{A}}{t}
    - \frac{3}{2}\OmegaK x\pder{\vec{A}}{y}
    = \frac{3}{2}\OmegaK A_y\unitvec_x
    + \vec{u}\times\left(\vec{B} + \vec{B}_\mathrm{ext}\right)
    - \mu_0\eta(z)\vec{J}\label{E:mhd_ind}
\end{equation}
\citep{BN95}.
The first term on the right-hand side is the magnetic stretching due to the background shear, and the last term is the Ohmic resistance with $\eta(z)$ being the magnetic diffusivity as a function of vertical position.
Finally, the dynamical part of the magnetic field is obtained by $\vec{B} = \curl{\vec{A}}$.

\begin{figure}
\plotone{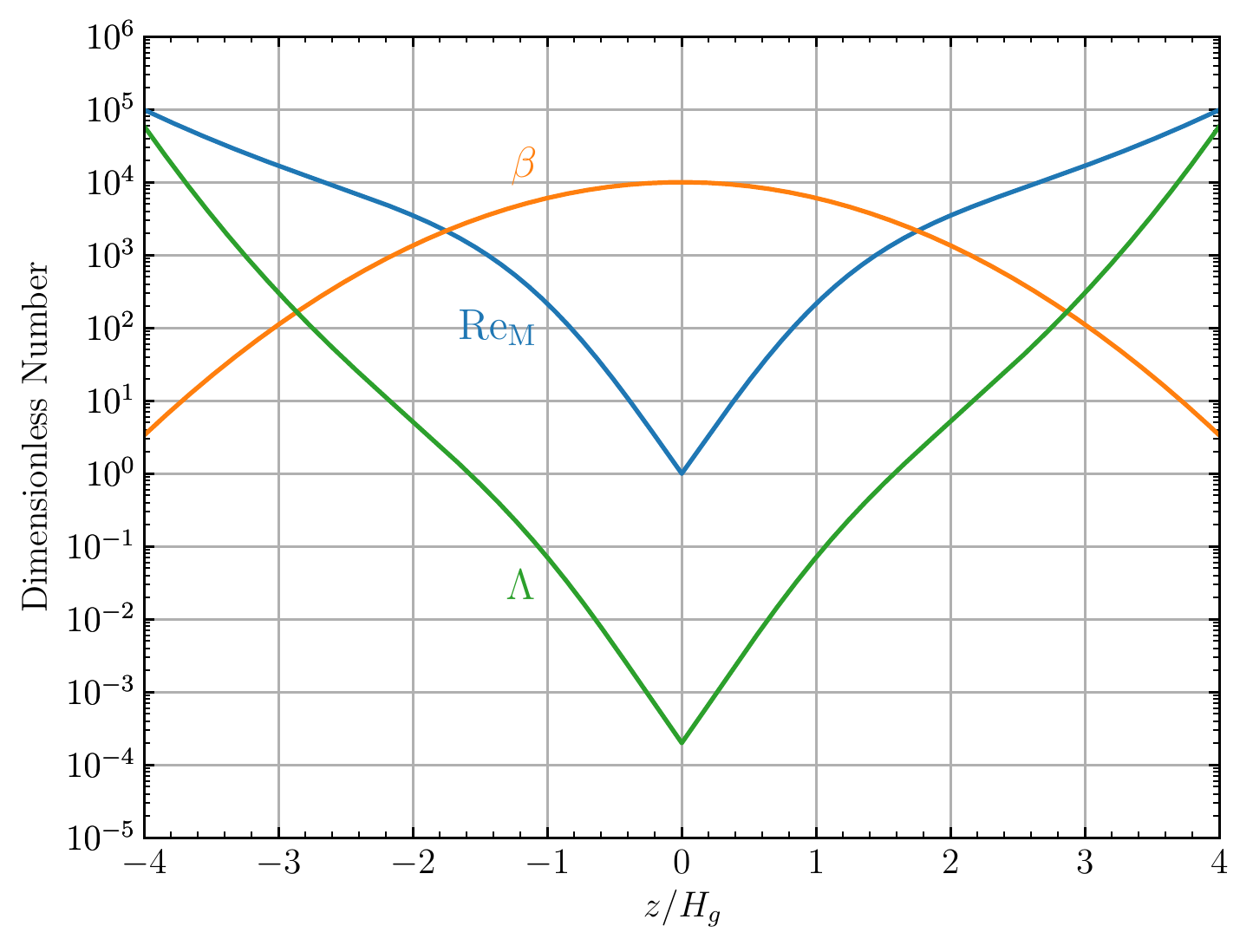}
\caption{Initial vertical profiles of the plasma $\beta$, the magnetic Reynolds number $\ReM$, and the Elsasser number $\Lambda$.
These dimensionless numbers are defined by Equations~\eqref{E:beta}, \eqref{E:ReM}, and~\eqref{E:Lambda}, respectively.
The condition $\Lambda \lesssim 1$ shows that the Ohmic resistance in our dead-zone models is initially effective in dissipating the MRI up to $|z| \simeq 1.6H_g$. \label{F:ivp}}
\end{figure}

Because we include the Ohmic resistance as our only non-ideal MHD term, we are modelling a layered accretion disk around a protostar \citep{cG96}, and for this purpose we adopt the resistivity profile of \citet{FS03}.
Their Ohmic diffusivity as a function of vertical position reads
\begin{equation}
\eta(z) =
\eta_0\exp\left[-\frac{z^2}{4H_g^2} +
                7.5\,\mathrm{erfc}\left(\frac{|z|}{\sqrt{2}H_g}\right)
          \right],\label{E:fs03}
\end{equation}
where $\eta_0$ is a constant coefficient and $H_g \equiv c_s / \OmegaK$ is the vertical scale height of the gas.
To create a dead zone of significant size, we set $\eta_0$ such that the magnetic Reynolds number
\begin{equation}
\ReM \equiv \frac{c_s^2}{\eta(z)\OmegaK}\label{E:ReM}
\end{equation}
is unity in the mid-plane \citep[cf.,][]{OMM07}, and Figure~\ref{F:ivp} shows the profile of our $\ReM$.
This resistivity profile was obtained by considering cosmic rays or X-rays as the only source of ionization with an assumed decay length in the vertical direction and ignoring the effects of solid grains.
We acknowledge that detailed calculation of the ionization structure in protoplanetary disks is still under active research \citep[see, e.g.,][and references therein]{TF14}; we note that Ohmic dissipation may dominate over ambipolar diffusion in the inner region ($\lesssim$3\,au) in a typical protoplanetary disk, and the effect of Hall drift in this region remains unclear \citep{xB17,BLF17}.
Nevertheless, as noted by \citet{OH11}, the gas dynamics inside a layered accretion disk predominantly depend on the sizes of the dead zone and the active layer and are rather insensitive to the details of the resistivity profile within the dead zone (see the discussion in the end of Section~\ref{SS:vf}, however).
Therefore, our use of Equation~\eqref{E:fs03} remains heuristic.

\subsubsection{Solid Particles} \label{SSS:par}

For the solid materials in the protoplanetary disk, we adopt the approach of Lagrangian super-particles.
Each super-particle has its own position $\vec{x}_p = (x_p, y_p, z_p)$ and velocity $\vec{v} = (v_x, v_y, v_z)$, where $\vec{v}$ is measured with respect to the background Keplerian shear $-3\OmegaK x_p\unitvec_y / 2$, and the super-particle represents numerous identical physical solid particles.
The equations of motion for each super-particle are then given by
\begin{align}
\tder{\vec{x}_p}{t}
    &= -\frac{3}{2}\OmegaK x_p\unitvec_y
    + \vec{v},\label{E:par_vel}\\
\tder{\vec{v}}{t}
    &= \left(2\OmegaK v_y\unitvec_x
        - \frac{1}{2}\OmegaK v_x\unitvec_y
        - \OmegaK^2 z_p\unitvec_z\right)
    + \frac{\vec{\tilde{u}} - \vec{v}}{t_s}.\label{E:par_acc}
\end{align}
The terms in parentheses in Equation~\eqref{E:par_acc} are parallel to those in Equation~\eqref{E:mhd_mom}.
The last term in Equation~\eqref{E:par_acc} is from the resultant drag force on the super-particle exerted by the surrounding gas, where $\tilde{\vec{u}}$ is the effective gas velocity experienced by the particle \citep{YJ07,YJ16}.

For simplicity, we assume that the stopping time $t_s$ is constant and the same for all the solid particles.
In the Epstein drag regime, where a particle is smaller than the mean free path of its surrounding gas and its velocity relative to the gas is much smaller than the speed of sound $c_s$, $t_s = \rho_s a / \rho_g c_s$, in which $\rho_s$ and $a$ are the material density and radius of the particle, respectively.
As is shown in Sections~\ref{S:ssgas} and \ref{S:sspar}, the perturbation in the gas density in the mid-plane is about 10\% and the scale height of the particle layer is about 0.2--0.3\,$H_g$.
Thus, the gas density the particles experience can be considered roughly constant and the perturbation in the gas density can be treated as a higher-order effect.
Therefore, our assumption of a constant $t_s$ can be translated into solid particles of approximately the same size.

In this work, we focus on solid particles with dimensionless stopping time $\tau_s \equiv t_s\OmegaK = 0.1$.
For the minimum mass solar nebula (\citealt{sW77b,cH81}), this corresponds to $\sim$dm-sized compact particles in the inner disk ($\lesssim$5\,au) and mm--cm sizes in the outer disk ($\gtrsim$5\,au) \citep[see, e.g.,][]{JB14}.
Particle coagulation limited by radial drift indeed reaches $\tau_s \sim 0.1$, as likely occurs outside the ice line \citep{BKE12}.

\subsection{Initial and Boundary Conditions} \label{SS:ibc}

The gas is initiated in hydrostatic equilibrium.
The initial density profile of the gas is then
\begin{equation}
\rho_{g,0}(z) = \rho_0\exp\left(-\frac{z^2}{2H_g^2}\right),\label{E:rhog0}
\end{equation}
where $\rho_0$ is the initial density of the gas in the mid-plane.
In order to seed the MRI, we apply an initial isotropic random perturbation of magnitude $10^{-3}c_s$ to the gas velocity $\vec{u}$.
We set the magnetic vector potential $\vec{A}$ to be initially zero and hence $\vec{B} = \vec{0}$.
On the other hand, we assign the magnitude of the external magnetic field $B_\mathrm{ext}$ such that the plasma
\begin{equation}
\beta
\equiv \frac{\rho_g c_s^2}{|\vec{B} + \Bext|^2 / 2\mu_0}
= \frac{2c_s^2}{v_A^2}\label{E:beta}
\end{equation}
is initially $\beta_0 = 10^4$ in the mid-plane, where $v_A \equiv |\vec{B} + \Bext| / \sqrt{\mu_0\rho_g}$ is the Alfv\'{e}n speed.
In other words, $B_\mathrm{ext} \simeq 0.014~c_s\sqrt{\mu_0\rho_0}$, where the units for the magnetic field are given by
\begin{align}
[B]
&= c_s\sqrt{\mu_0\rho_0}\nonumber\\
&= \left(2.8\times10^{-4}\,\textrm{T}\right) \times\nonumber\\
&\qquad\left(\frac{c_s}{8\times10^2~\textrm{m\,s}^{-1}}\right)
   \left(\frac{\rho_0}{10^{-7}~\textrm{kg\,m}^{-3}}\right)^{1/2}.
\label{E:bunits}
\end{align}
This places the critical wavelength of the ideal MRI near the mid-plane at $\sim$0.026$H_g$ \citep{BH91}.
With Equation~\eqref{E:rhog0}, the initial vertical profiles of plasma $\beta$ and the Elsasser number $\Lambda$, which is defined by
\begin{equation}
\Lambda
\equiv \frac{v_A^2}{\eta(z)\OmegaK}
= \frac{2c_s^2}{\beta\eta(z)\OmegaK}
= \frac{2\ReM}{\beta},\label{E:Lambda}
\end{equation}
are shown in Figure~\ref{F:ivp}.
Given that the condition $\Lambda \sim 1$ determines the upper boundary where the Ohmic resistance becomes effective in dissipating the MRI \citep{SM99,OH11}, Figure~\ref{F:ivp} indicates that the initial extent of our dead zone covers the region $|z| \lesssim 1.6H_g$.

We adopt a computational domain of $4H_g \times 8H_g \times 8H_g$ in the radial, azimuthal, and vertical directions.
In the horizontal dimensions, we use the standard sheared periodic boundary conditions \citep{BN95,HGB95}.
In the vertical dimension, we apply zero-order extrapolations, i.e.,
\begin{align}
f(t,x,y,z) &= f(t,x,y,z_b)\textrm{ for }z < z_b,\\
f(t,x,y,z) &= f(t,x,y,z_t)\textrm{ for }z > z_t,
\end{align}
where $f$ is any dynamical field except the gas density $\rho_g$, and $z_b$ and $z_t$ are the vertical coordinates of the last active grid cells in the bottom and the top, respectively.
For the gas density field $\rho_g$, we adopt the same boundary conditions as in \citet{SHB11}.
These boundary conditions instead extrapolate the ratio of the gas density to the initial equilibrium density profile $\rho_{g,0}$ (Equation~\eqref{E:rhog0}):
\begin{align}
\rho_g(t,x,y,z) &= \frac{\rho_g(t,x,y,z_b)}{\rho_{g,0}(z_b)}\rho_{g,0}(z)
\textrm{ for }z < z_b,\\
\rho_g(t,x,y,z) &= \frac{\rho_g(t,x,y,z_t)}{\rho_{g,0}(z_t)}\rho_{g,0}(z)
\textrm{ for }z > z_t.
\end{align}
We note that these vertical boundary conditions practically achieve nonreflecting boundary conditions with respect to the initial density stratification, and these boundary conditions are equivalent to the zero-order extrapolations applied to the hyperbolic system formulated in \citet{YJ14}, in which $\rho_{g,0}(z)$ is factored out.\footnote{For more discussion on nonreflecting boundary conditions, see, e.g., \citet[Chapter~7]{rL02}.}

We allow the system of gas to evolve for about 10--20\,$P$, where $P \equiv 2\pi / \OmegaK$ is the local orbital period, so that it reaches a statistically steady state of MHD turbulence before initiating the solid particles.
First, we activate the background radial acceleration term $2\Pi c_s\OmegaK\unitvec_x$ to the gas with $\Pi = 0.05$, a typical value in the inner region of a solar nebula \citep{BS10,BJ15}.
We then allocate as many Lagrangian particles as the total number of grid cells and randomly distribute them in a vertical Gaussian distribution with a scale height of 0.2$H_g$ or 0.3$H_g$.
(The exact choice of the initial scale height does not noticeably affect the saturation stage of the particle-gas dynamics.)
Assuming that all the particles have the same mass and combine to have a solid abundance $Z \equiv \Sigma_{p,0} / \Sigma_{g,0}$, where $\Sigma_{p,0}$ and $\Sigma_{g,0} = \sqrt{2\pi}\rho_0 H_g$ are the initial column densities of the solids and the gas, respectively, the mass of each particle is determined \citep{YJ14,YJC17}. 
Finally, to obtain an initial local dynamical balance, we add the Nakagawa--Sekiya--Hayashi \citeyearpar{NSH86} solutions for the equilibrium velocities to both the gas (on top of the saturated turbulence) and the particles; the initial vertical velocity of the particles are set zero.

The particles also observe the sheared periodic boundary conditions \citep{YJ07,YJ16}.
The vertical boundary conditions for the particles are set periodic, although none of the particles move close to the vertical boundaries in practice.

\subsection{Numerical Methods}

We use the \pc\ \citep{BD02} to numerically integrate the system of Equations~\eqref{E:mhd_cont}, \eqref{E:mhd_mom}, \eqref{E:mhd_ind}, \eqref{E:par_vel}, and \eqref{E:par_acc}.
The \pc\ is a cache efficient, massively parallelized code suitable for MHD turbulence on an Eulerian grid coupled with Lagrangian particles.
It uses sixth-order finite differences to calculate all the spatial derivatives on the grid to achieve high fidelity at high wavenumbers, and it employs third-order Runge--Kutta integration in time \citep{aB03}.
Sixth-order hyper-diffusion operators on each dynamical field are required to stabilize the scheme.
For these operators, we fix the mesh Reynolds number to target numerical damping near the Nyquist frequency while preserving the power over a wide dynamical range \citep{YK12}.
To capture the shocks in the flow, artificial diffusion operators on each dynamical field are also needed.
Instead of using a shock diffusion coefficient of von Neumann type, as commonly employed in the \pc{} \citep{HBM04}, we use the HLLE solution to estimate the maximum local shock speed and in turn use it to compute the diffusion coefficient, which proves to be superior in high-altitude regions (C.-C.\ Yang, in preparation).

To relieve the Courant condition limited by the background shear and reduce the associated radially dependent numerical diffusion, we adopt the algorithm of shear advection by interpolation developed by \citet{JYK09}.
Instead of using Fourier interpolations, however, we use B-spline interpolations.
The reason is that whenever shocks are present, Fourier interpolations suffer from the Gibbs phenomenon and tend to increase the total variation of the field, leading to numerical instability.
On the other hand, B-splines have the desirable property of total-variation-diminishing and can be designed to achieve an accuracy of arbitrary order.
We choose sixth-order B-splines to match the accuracy of the spatial derivatives used in the \pc.
For more information on B-splines, readers are referred to \citet{cB78}.

Another numerical difficulty comes from the resistance term in Equation~\eqref{E:mhd_ind}.
Given the vertical profile of the magnetic Reynolds number shown in Figure~\ref{F:ivp}, this term is particularly stiff near the mid-plane.
We describe our algorithm to integrate this term in Appendix~\ref{S:isor}.

With net vertical magnetic field as in our models, the MHD turbulence at high altitudes tends to drive disk winds, leading to gradual loss of disk mass \citep{SI09}.
However, the mass loss rate is numerically sensitive to the vertical dimension of the shearing box \citep{BS13,FL13} and our limited computational domain would induce an artificially large mass loss rate.
In this work, therefore, we enforce mass conservation and seek a statistically steady state of the MHD turbulence.
At each time step, we apply a constant factor to the gas density field to maintain a constant total gas mass while adjusting the gas velocity in each cell so that the momentum of the gas in the cell remains the same.
This approach to achieve a quasi-steady state solution is commonly adopted in the literature \citep[e.g.,][]{gO12,BS13,LKF14}.

Finally, the equations of motion for the super-particles (Equations~\eqref{E:par_vel} and \eqref{E:par_acc}) are integrated synchronously  with the Eulerian gas using the same Runge--Kutta steps.
To achieve high accuracy in the coupling between the Eulerian gas and the Lagrangian particles, we use the standard Triangular-Shaped-Cloud scheme for the particle-mesh interpolation and assignment \citep{HE88}.
We adopt the algorithm developed by \citet{YJ07} for the mutual drag force to ensure momentum conservation.
Given that our vertical dimension is large compared to the scale height of the particle layer, we employ the algorithm of particle block domain decomposition designed by \citet{JKH11} to obtain better load balancing in parallel computing.
As a side note, we recently developed a new numerical algorithm for the mutual drag force in \citet{YJ16}, which relieves the time-step constraint limited by small stopping time and/or high local solid-to-gas density ratio, but this algorithm has yet to be implemented with the particle block domain decomposition.
Nevertheless, the stopping time we investigate in this work is relatively large and hence the major bottleneck in computing efficiency is the load balance in the distribution of particles instead of the time steps.
Therefore, we prefer the particle block domain decomposition to the new integration scheme for the mutual drag force.

\section{QUASI-STEADY-STATE PROPERTIES\\OF THE GAS} \label{S:ssgas}

In this section, we focus on several diagnostics of the gas in the statistically steady state of the gas flow in our various models without back reaction of solid particles.
These diagnostics establish a base for comparison with similar MHD calculations in the literature \citep[e.g.,][]{OH11,ZSB15,RL18}.
More importantly, they help us understand the dynamical response of the particles to the gas motions.
We consider disks with and without Ohmic resistivity (Equation~\eqref{E:fs03}) so that we can compare the particle-gas dynamics between a dead zone and ideal MHD.
The ideal-MHD models have a resolution of 16 points per gas scale height $H_g$, while the dead-zone models have a resolution of 16 or 32 points per $H_g$.

\subsection{Mean Vertical Profiles} \label{SS:mvp}

To obtain the mean vertical profile of a property $f$, we horizontally average it at each vertical position $z$ at any given instant, yielding $\langle f\rangle_z$, and then time average the results.
Since most of our diagnostics are positive definite quantities and cover several orders of magnitude, we conduct the time average in logarithmic space as
\begin{equation} \label{E:tavg}
\overline{\langle f\rangle_z} \equiv
\exp\left[\frac{1}{t_2 - t_1}\int_{t_1}^{t_2}
          \ln \langle f\rangle_z\mathrm{d}t\right],
\end{equation}
where $t_1$ and $t_2$ are the integration limits in time.
The cadence of the snapshots is less than 0.1$P$, and we choose to integrate for $t_2 - t_1 = 100P$, where $P$ is the orbital period.

\begin{figure*}
\gridline{\fig{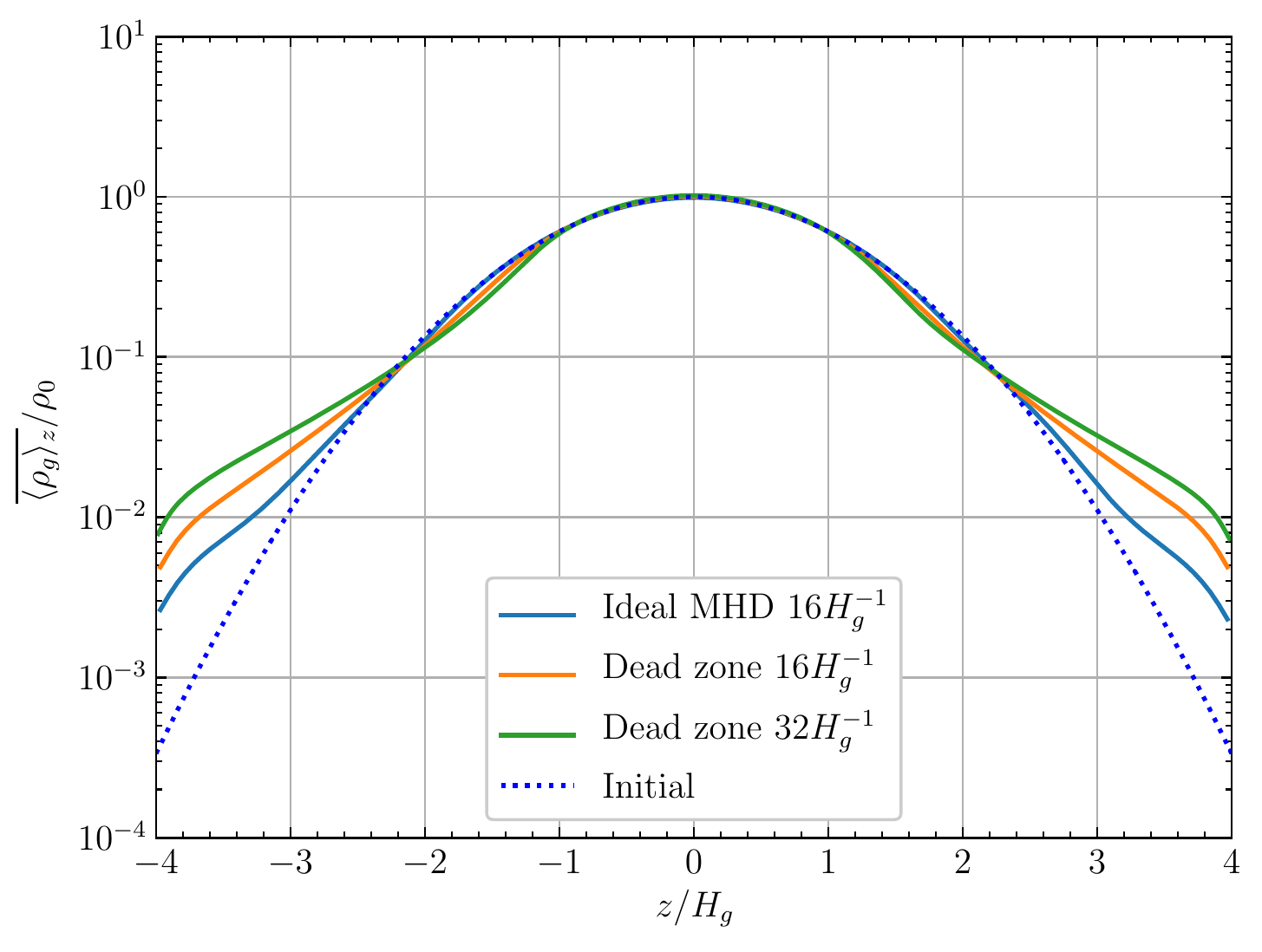}{0.333\textwidth}{(a)~Gas density}
          \fig{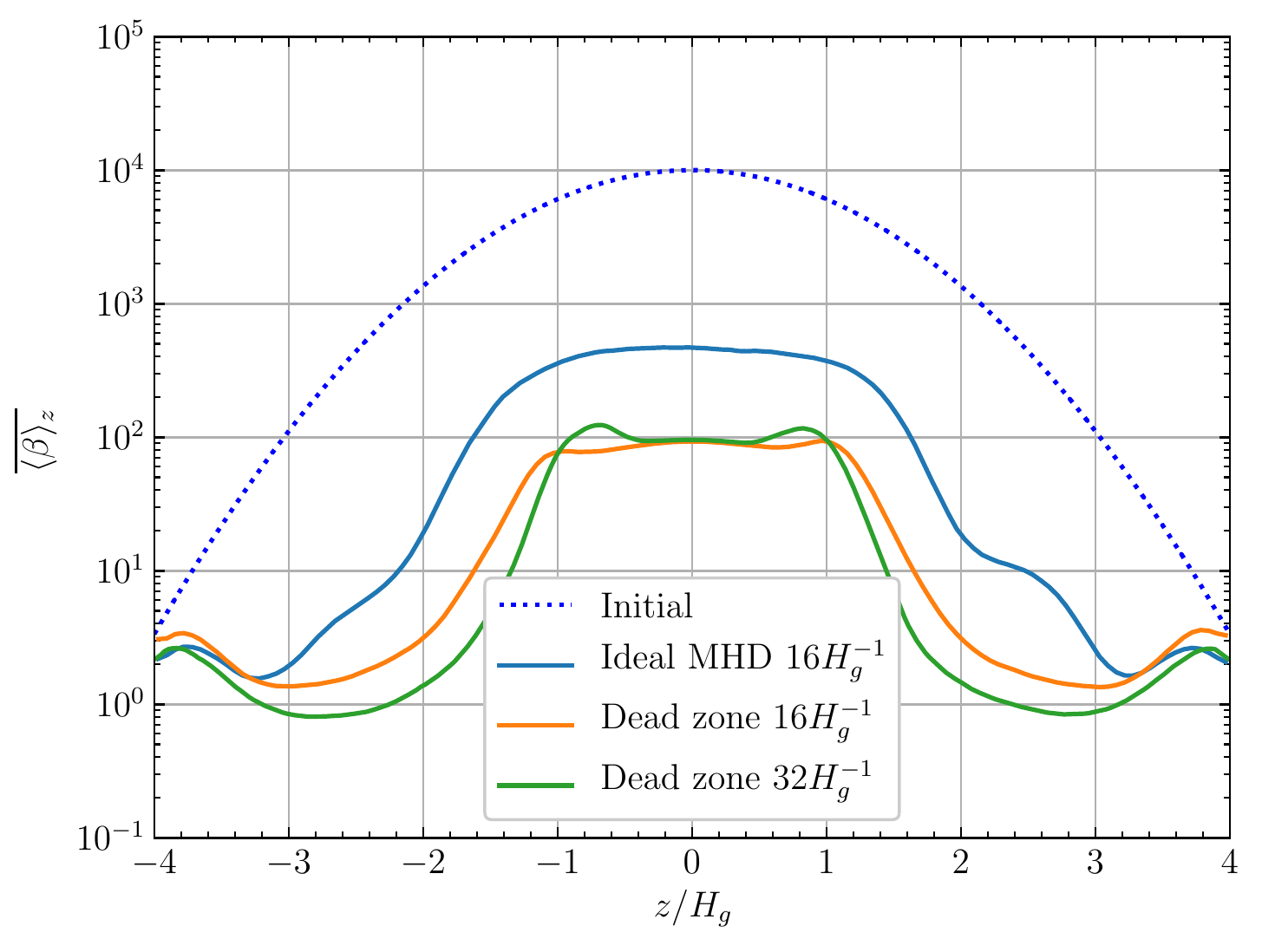}{0.333\textwidth}{(b)~Plasma $\beta$}}
\gridline{\fig{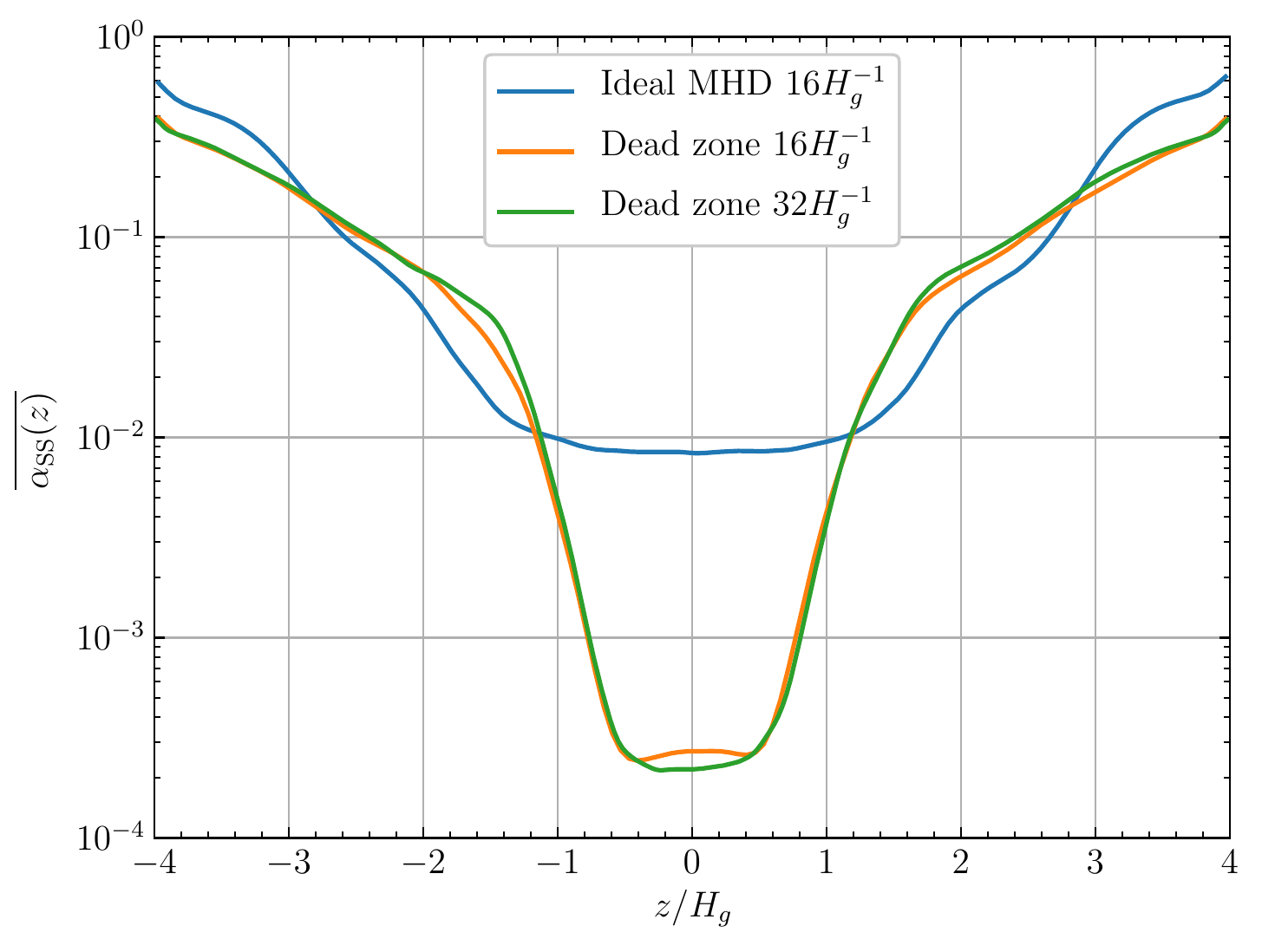}{0.333\textwidth}{(c)~\ssa{} stress parameter}
          \fig{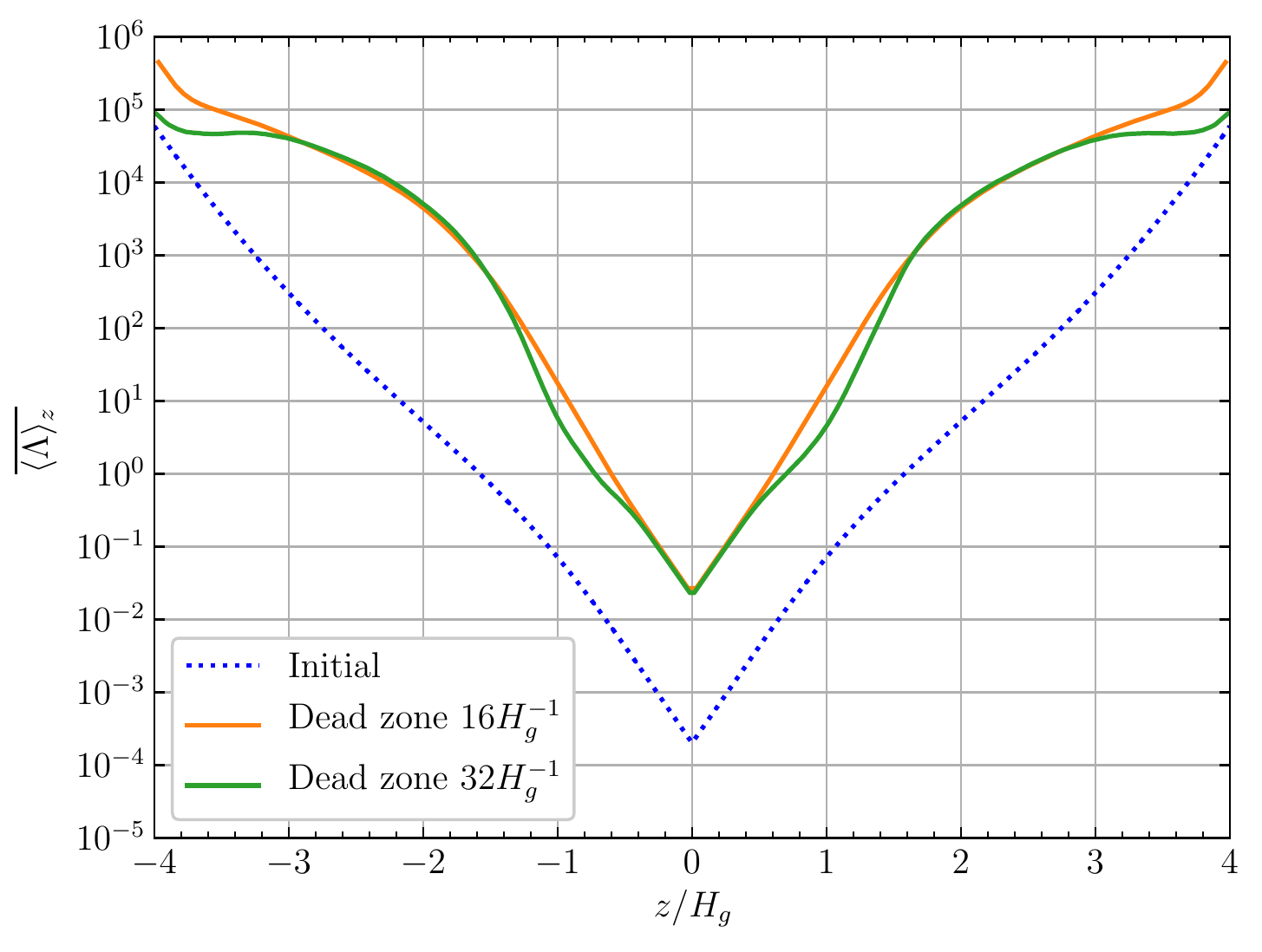}{0.333\textwidth}{(d)~Elsasser number}}
\gridline{\fig{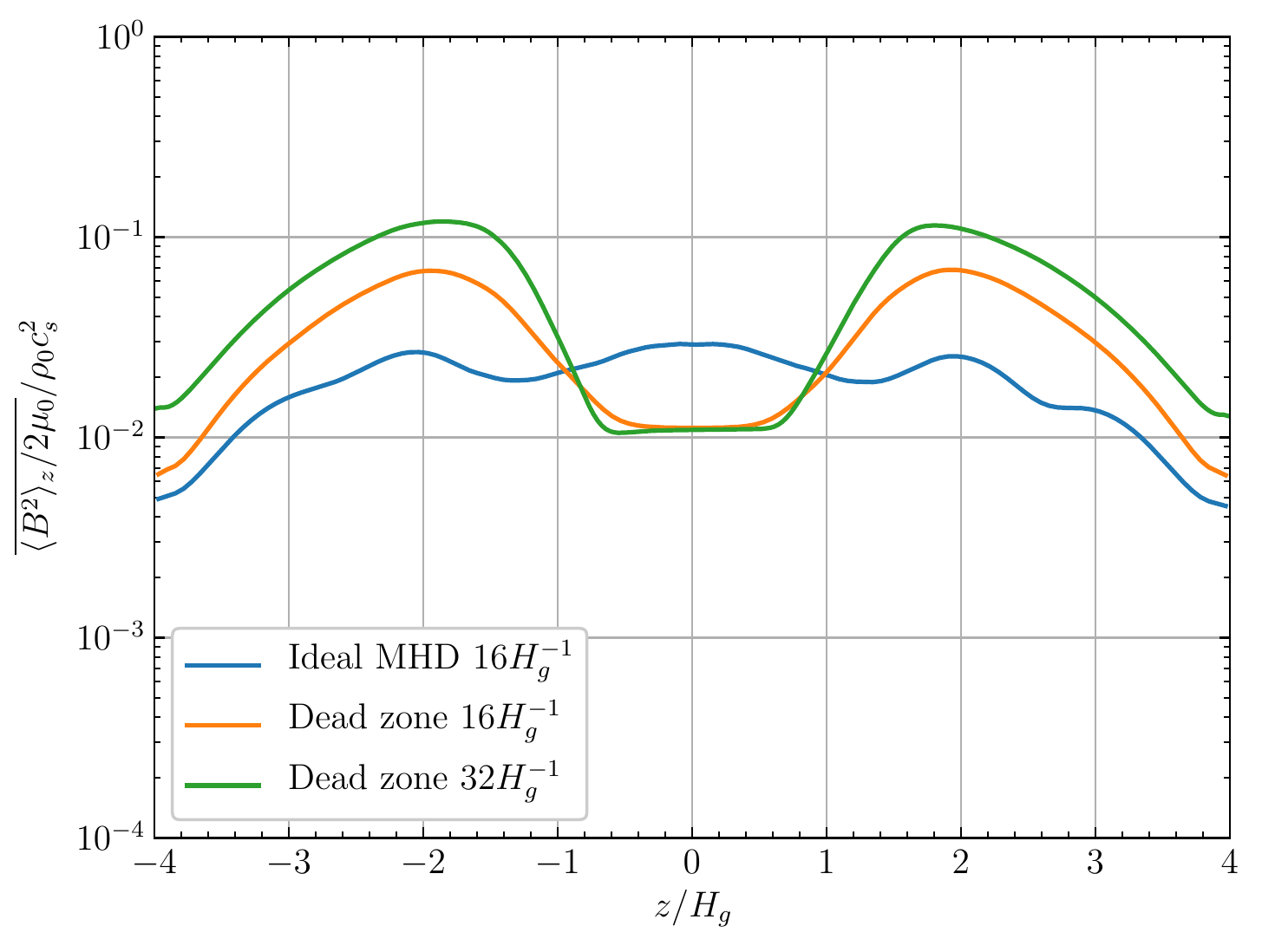}{0.333\textwidth}{(e)~Magnetic energy density}
          \fig{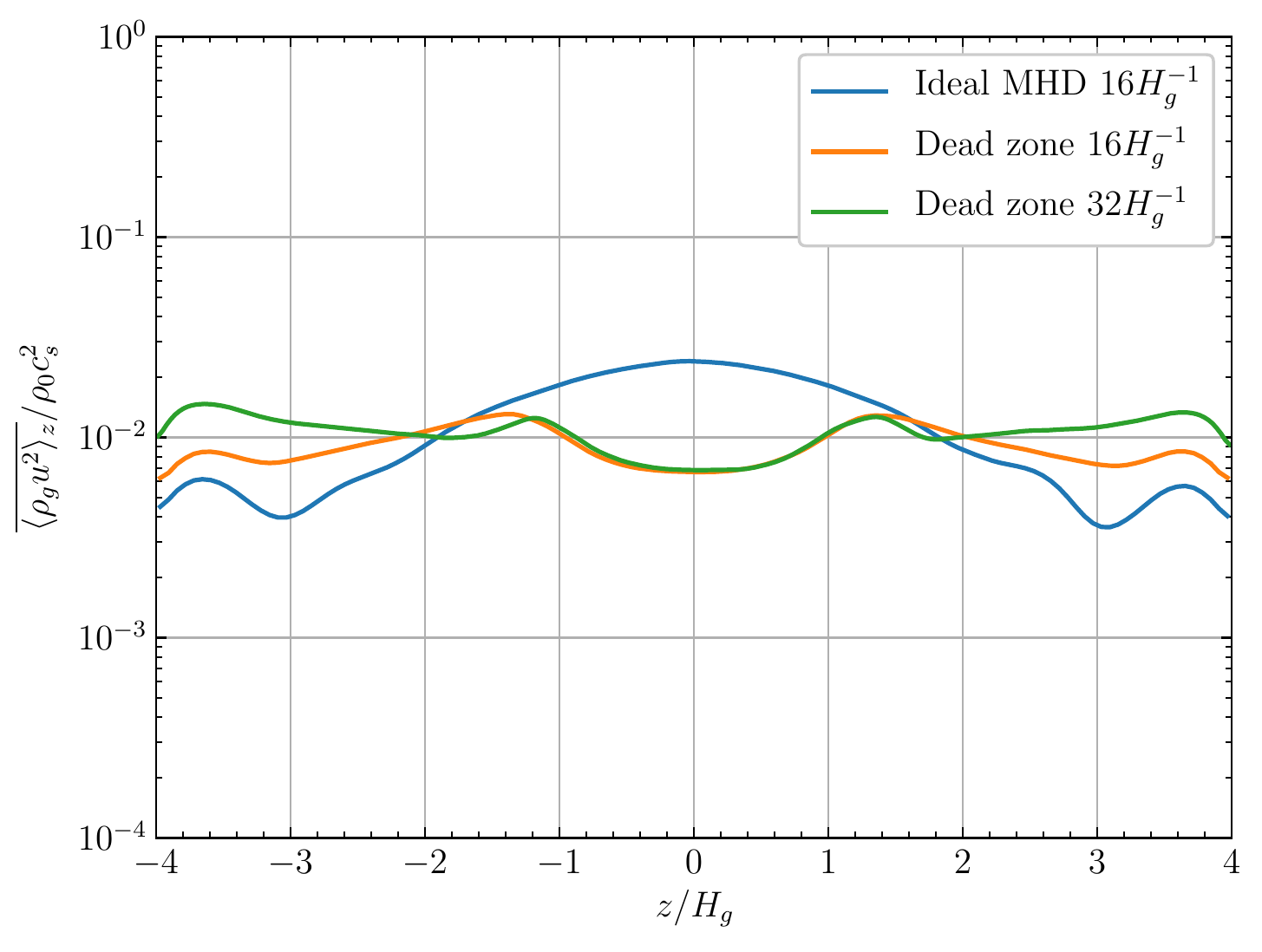}{0.333\textwidth}{(f)~Kinetic energy density}
          \fig{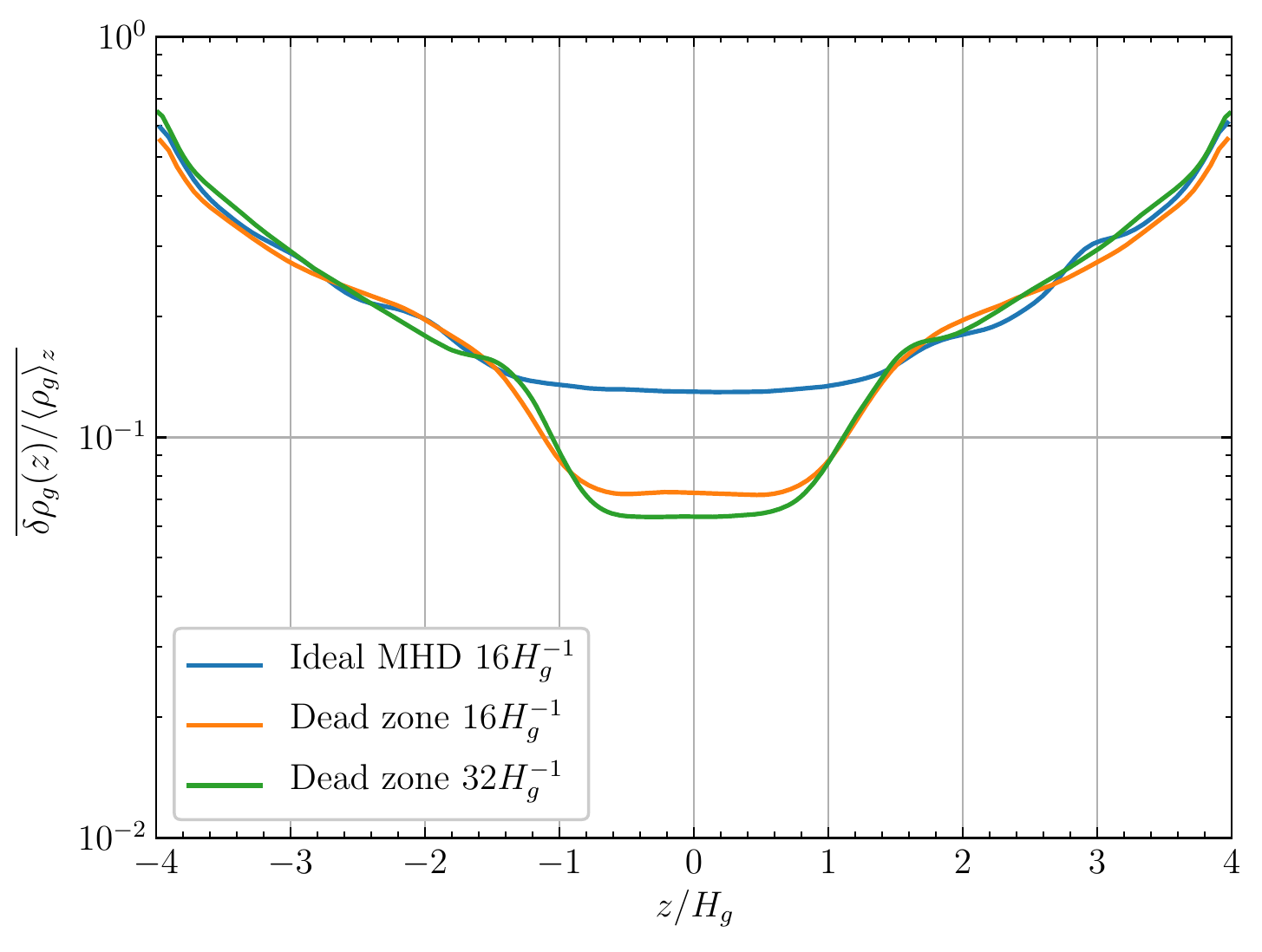}{0.333\textwidth}{(g)~Gas density fluctuation}}
\caption{Mean vertical profiles of gas properties in the saturated state of the MHD turbulence in various models without the back reaction of the solid particles.
    All the properties are horizontally averaged at each snapshot with a cadence of less than 0.1$P$ and then time averaged over a period of 100$P$.
    The solid lines of different colors represent different models.
    The dotted lines in panels~(a), (b), and~(d) denote the initial equilibrium profile.\label{F:vpgas}}
\end{figure*}

Figure~\ref{F:vpgas}a shows the mean vertical profiles of the gas density along with the initial hydrostatic equilibrium profile (Equation~\eqref{E:rhog0}).
The mean profiles for all our models closely follow the initial profile up to $z \simeq \pm2.2H_g$, with a slight decrement for $1 \lesssim |z / H_g| \lesssim 2.2$ in our dead-zone models. 
For high altitudes $|z| \gtrsim 2.2H_g$, on the other hand, a significant increase in gas density compared to hydrostatic equilibrium is observed in all our models.
This may be understood by noting the increasing support of magnetic pressure towards higher altitudes \citep{TCS10,OH11,BS13}, as shown by the plasma $\beta$ in Figure~\ref{F:vpgas}b, where $\beta$ is the ratio of the thermal pressure to the magnetic pressure (Equation~\eqref{E:beta}).
The value of $\beta$ is appreciably less than 10 for $|z| \gtrsim 2.2H_g$, and thus the magnetic pressure is of the same order of magnitude as the thermal pressure.
This effect of extra pressure support and denser gas at high altitudes is stronger in our dead-zone models than in our ideal-MHD models.
We note, however, that the vertical profile of the gas density fluctuates significantly over time at these altitudes, driven by intermittent launch of a large-scale, outflowing disk wind.

The next quantity of interest is the \ssa{} (\citeyear{SS73}) stress parameter, which is a dimensionless measure of the turbulent viscosity.
Following \cite{aB98}, we denote the parameter by $\ass$ and calculate it as a function of vertical position by
\begin{equation} \label{E:ass}
\ass(z) \equiv
\frac{\langle\rho_g u_x u_y\rangle_z - \langle B_x B_y\rangle_z / \mu_0}
     {3\langle\rho_g\rangle_z c_s^2 / \sqrt{2}},
\end{equation}
where the first and the second terms in the numerator are the Reynolds and Maxwell stresses, which are normalized by the mean pressure at the given $z$ and scaled by the Keplerian shear.
The resulting time-averaged vertical profiles for our various models are shown in Figure~\ref{F:vpgas}c.

The existence of a dead zone in our models with Ohmic resistance is apparent by comparing the $\ass$ profiles.
The $\ass$ stress in the mid-plane of our ideal-MHD model is on the order of $10^{-2}$, which is consistent with those measured in previous works \citep{YMM09,YMM12,BS13}, considering our imposed vertical magnetic field with $\beta_0 = 10^4$ (Section~\ref{SS:ibc}).
On the other hand, the turbulent stress near the mid-plane of our dead-zone models is significantly less, with $\ass \sim 2\times10^{-4}$.
This is more than an order of magnitude smaller than in our ideal-MHD model.
The turbulent stress is relatively indistinguishable between the dead-zone and ideal-MHD models for $|z| \gtrsim 1.2H_g$, indicating the extent of the dead zone is roughly up to that altitude.

Closely related to the mean vertical profiles of the $\ass$ stress in our dead-zone models are those of the Elsasser number $\Lambda$ (Equation~\eqref{E:Lambda}), as shown in Figure~\ref{F:vpgas}d.
The Elsasser number in the saturated state of turbulence is significantly higher than in the initial conditions, due to the much increased magnetic activity throughout the computational domain.
This shifts the critical location of $\Lambda \sim 1$ from $z \simeq \pm 1.6H_g$ to $z \simeq \pm 0.6$--0.7$H_g$, coincident with the extent of the flat bottom in the mean $\ass$ profiles observed in Figure~\ref{F:vpgas}c.

The effect of Ohmic resistance can also be seen in the vertical profiles of magnetic and kinetic energy densities, as shown in Figures~\ref{F:vpgas}e and~\ref{F:vpgas}f, respectively.
In comparison to the ideal-MHD model, both energy densities near the mid-plane in the dead-zone models are depressed by about a factor of three.
This reduction is appreciably smaller than the reduction in the shear stress as measured by $\ass$, indicating that shear stress and energy density are not necessarily linearly related in the non-ideal MHD flow in the dead zone.
This observation has important consequences in the study of particle-gas dynamics in the dead zone, as discussed in Section~\ref{S:sspar}.
We note also that the energy densities in the dead-zone models exceed those of the ideal-MHD model at high altitudes.
A layered accretion disk drives more activity in the transition region between magnetically active and dead zones.

Finally, Figure~\ref{F:vpgas}f shows the mean vertical profiles of the relative density fluctuation $\delta\rho_g(z) / \ha{\rho_g}{z}$ in our models, where $\delta\rho_g(z) \equiv (\ha{\rho_g^2}{z} - \ha{\rho_g}{z}^2)^{1/2}$.
The perturbation near the mid-plane of the ideal-MHD model is about 13\%, while the dead zone still has a perturbation of about 6--7\%.
Near the vertical boundary $z \sim \pm4H_g$, on the other hand, all the models show density fluctuations as high as $\sim$60\%.

\begin{figure*}
\plotone{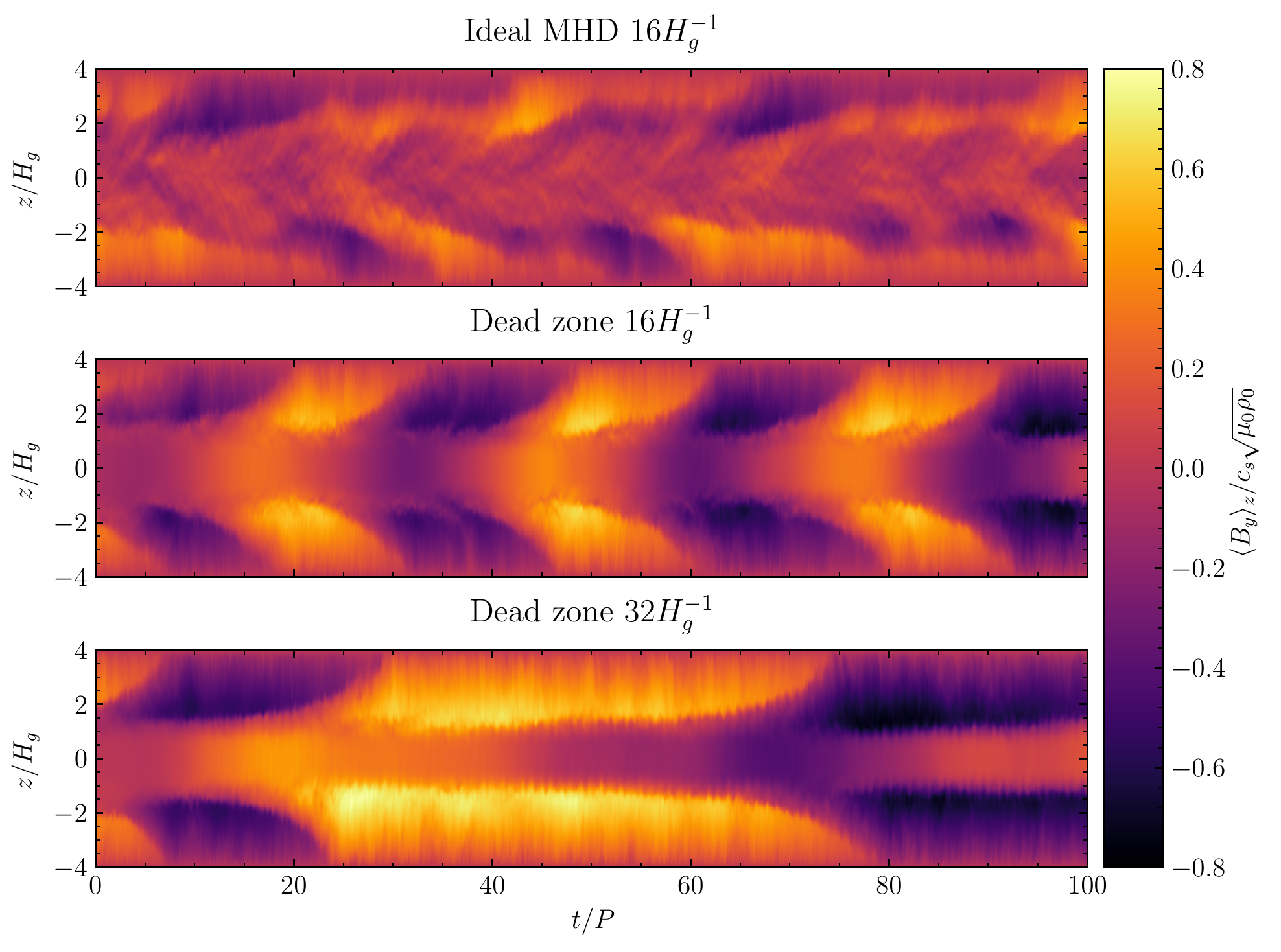}
\caption{Horizontally averaged azimuthal magnetic field as a function of time for various models.
The physical units for the magnetic field are given in Equation~\eqref{E:bunits}.
The characteristic ``butterfly pattern'' can be seen in these models.
See the discussion at the end of Section~\ref{SS:mvp} on the differences between the models.
\label{F:bymz}}
\end{figure*}

Figure~\ref{F:bymz} further demonstrates the evolution of the azimuthal magnetic field driven by the MRI in our models.
The ideal-MHD model and the surface layers of the dead-zone models show the characteristic ``butterfly pattern'' often reported in the literature, where azimuthal fields are generated near the mid-plane or the base of the surface layers, respectively, and then rise out of the mid-plane over time \citep[e.g.,][]{SH96,FS03}.
In our dead-zone models, we note that the frequency for the change of polarity in the butterfly pattern depends on resolution; the higher the resolution, the longer it takes to change polarity.
This behavior was also observed in the ideal stratified MHD models conducted by \cite{BS13}.
Moreover, we note that the dead zone is not necessarily magnetically ``dead''; significant azimuthal fields cyclicly occur near the mid-plane of our dead-zone models, a phenomenon unique to models with net vertical magnetic flux.
An understanding of these two effects is not yet complete, but is outside the scope of this paper, so we refer to the discussion by \cite{GNT11} and references therein.

\subsection{Velocity Fluctuations} \label{SS:vf}

Given that the particles and the gas interact via the drag force, the velocity fluctuations in the gas directly influence the dynamics of the particles.
We therefore measure two key statistical properties of the flow, the velocity dispersion and the correlation time of the random process, which then help us evaluate the diffusion of the particles in Section~\ref{S:sspar}.

We measure the velocity dispersion of the gas as a function of vertical position as follows.
First, at any given instant in time and vertical position $z$, we take the standard deviation of the gas velocity over all cells in the horizontal plane at $z$.
We denote the result by $\delta\vec{u}(z)$ and thus its components can be expressed as
\begin{equation}
\delta u_i(z) \equiv (\ha{u_i^2}{z} - \ha{u_i}{z}^2)^{1/2}.
\end{equation}
Then we take its time average using Equation~\eqref{E:tavg} with a duration of $t_2 - t_1 = 100P$, where $P$ is the orbital period.

\begin{figure}
\plotone{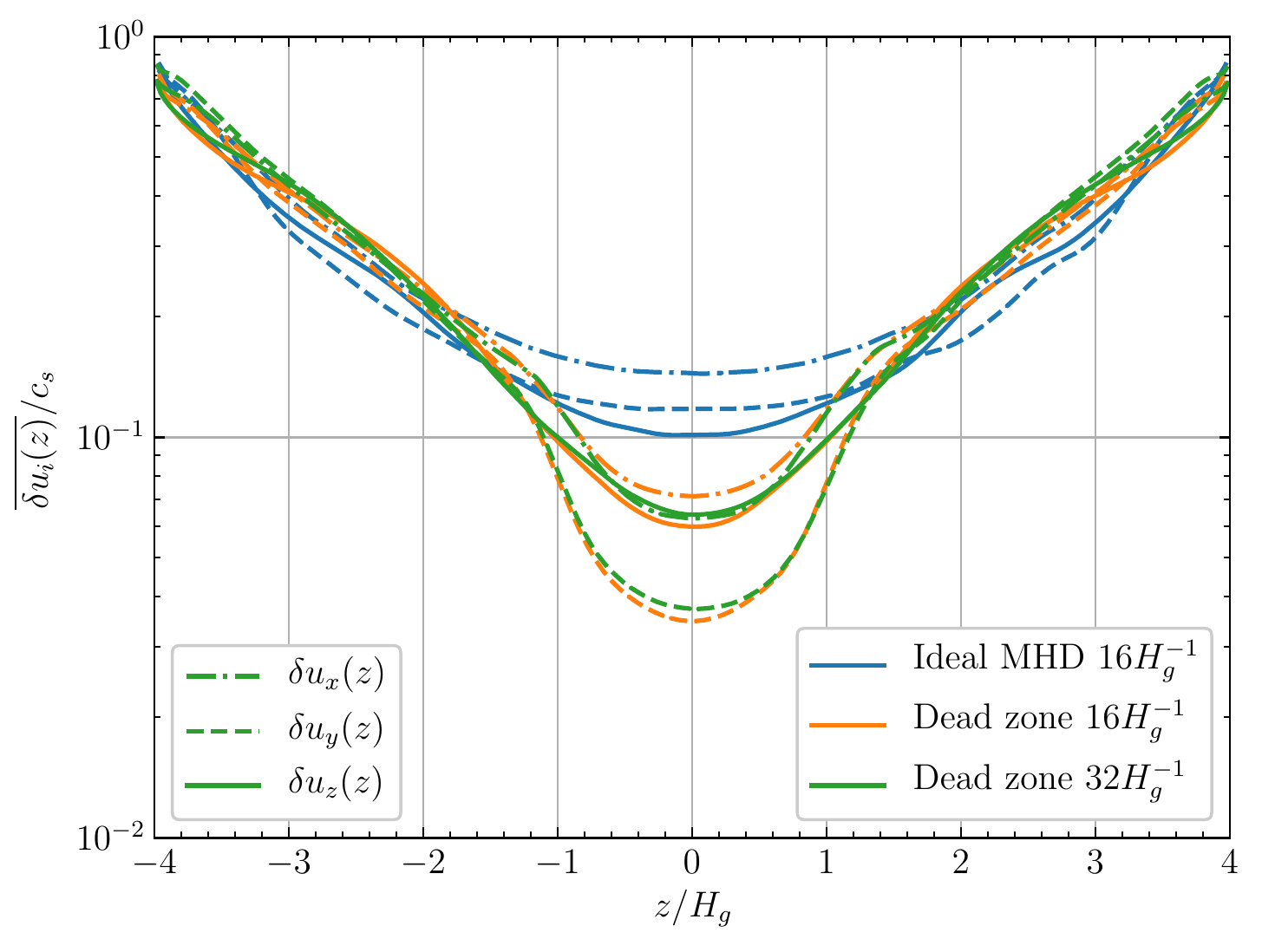}
\caption{Mean vertical profiles of the gas velocity dispersion in the saturated state, similar to Figure~\ref{F:vpgas}.
The dash-dotted, dashed, and solid lines denote the radial, azimuthal, and vertical components of the velocity dispersion, respectively.
Different colors represent different models.
The velocity dispersions are normalized by the speed of sound $c_s$.
The ideal-MHD model shows relatively isotropic turbulence, while inside the dead zone, the gas likely undergoes epicycle motions with azimuthal velocity dispersion only half of radial and vertical values.\label{F:vpdu}}
\end{figure}

\begin{deluxetable*}{lccccccccccc}
\tabletypesize{\scriptsize}
\tablecaption{Properties of the MHD flow in the mid-plane\label{T:du}}
\tablehead{\\
    &
    & \multicolumn{3}{c}{Velocity Dispersion}
    & \multicolumn{3}{c}{Correlation Time}
    & \multicolumn{3}{c}{Bulk Diffusion\tablenotemark{a}}
    & Shear Stress\tablenotemark{b}\\
\colhead{Model}
    & \colhead{Resolution}
    & \colhead{$\overline{\delta u_x(0)}$}
    & \colhead{$\overline{\delta u_y(0)}$}
    & \colhead{$\overline{\delta u_z(0)}$}
    & \colhead{$t_{c,x}(0)$}
    & \colhead{$t_{c,y}(0)$}
    & \colhead{$t_{c,z}(0)$}
    & \colhead{$\alpha_{g,x}(0)$}
    & \colhead{$\alpha_{g,y}(0)$}
    & \colhead{$\alpha_{g,z}(0)$}
    & \colhead{$\ass(0)$}\\
    & \colhead{($H_g^{-1}$)}
    & \colhead{($c_s$)}
    & \colhead{($c_s$)}
    & \colhead{($c_s$)}
    & \colhead{($P$)}
    & \colhead{($P$)}
    & \colhead{($P$)}
}

\startdata
Ideal MHD & 16 & 0.14(2) & 0.12(2) & 0.10(1) & 0.05 & 0.10 & 0.11
    & 0.0068 & 0.0083 & 0.0068    & 0.008(2)\phn\\
Dead Zone & 16 & 0.07(2) & 0.03(1) & 0.06(2) & 0.02 & 0.12 & 0.16
    & 0.0008 & 0.0009 & 0.0037 & 0.0003(7)\\
Dead Zone & 32 & 0.06(1) & 0.04(1) & 0.06(2) & 0.07 & 0.12 & 0.13
    & 0.0018 & 0.0010 & 0.0035 & 0.0002(9)\\
\enddata

\tablecomments{The standard deviation over time for each property is shown in parentheses.}
\tablenotetext{a}{Measured by the autocorrelation of the velocity fluctuations; see Equation~\eqref{E:gasdiff}.}
\tablenotetext{b}{Measured by the \ssa{} stress parameter; see Equation~\eqref{E:ass}.}

\end{deluxetable*}

The resulting vertical profiles of velocity dispersion for our models without the back reaction of the solid particles are shown in Figure~\ref{F:vpdu} (cf., Figures~4 and 14 of \citealt{FP06} and Figure~8 of \citealt{OH11}).
All the models demonstrate increasing velocity dispersion with height, reaching roughly the speed of sound near the vertical boundary of the computational domain.
The three components of the velocity dispersion for the ideal-MHD model show similar amplitudes at each height, indicating relatively isotropic turbulence across the whole domain.
On the other hand, inside the dead zone the velocity fluctuations are weaker, as expected, though only by a factor of a few, which is consistent with the profiles of the kinetic energy density measured in Figure~\ref{F:vpgas}f.
Moreover, $\delta u_x \sim 2\delta u_y \sim \delta u_z $ inside the dead zone, indicating that the gas likely undergoes epicyclic oscillations (see the discussion below, however).
In the active surface layer of the dead-zone models, the velocity dispersion becomes indistinguishable from that in the ideal-MHD model.
The measured values of the velocity dispersion in the mid-plane are listed in Table~\ref{T:du}.

We next measure the correlation time of the velocity fluctuations.
At each fixed point in space, we evaluate the autocorrelation of the gas velocity fluctuations over time, with each component of the autocorrelation denoted by
\begin{equation} \label{E:acf}
\mathscr{R}_i(t) \equiv
\int \left[u_i(\tau) - \overline{u_i}\right]
     \left[u_i(\tau+t) - \overline{u_i}\right]\mathrm{d}\tau,
\end{equation}
where $\overline{\vec{u}}$ is the mean velocity, which is estimated by taking  the time average of the gas velocity from $t_1$ to $t_2$:
\begin{equation}
\overline{u_i} \simeq
\frac{1}{t_2 - t_1}\int_{t_1}^{t_2} u_i(\tau)\mathrm{d}\tau.
\end{equation}
It is expected that the correlation time should not exceed the orbital timescale \citep{FP06,JKM06,OMM07,YMM09,YMM12}, and hence we use $t_2 - t_1 = 10P$ with a high cadence of $0.01P$ when recording the data for this purpose.
We take the horizontal average of Equation~\eqref{E:acf} to obtain a good ensemble average of the autocorrelation as a function of vertical position $z$.

\begin{figure}
\plotone{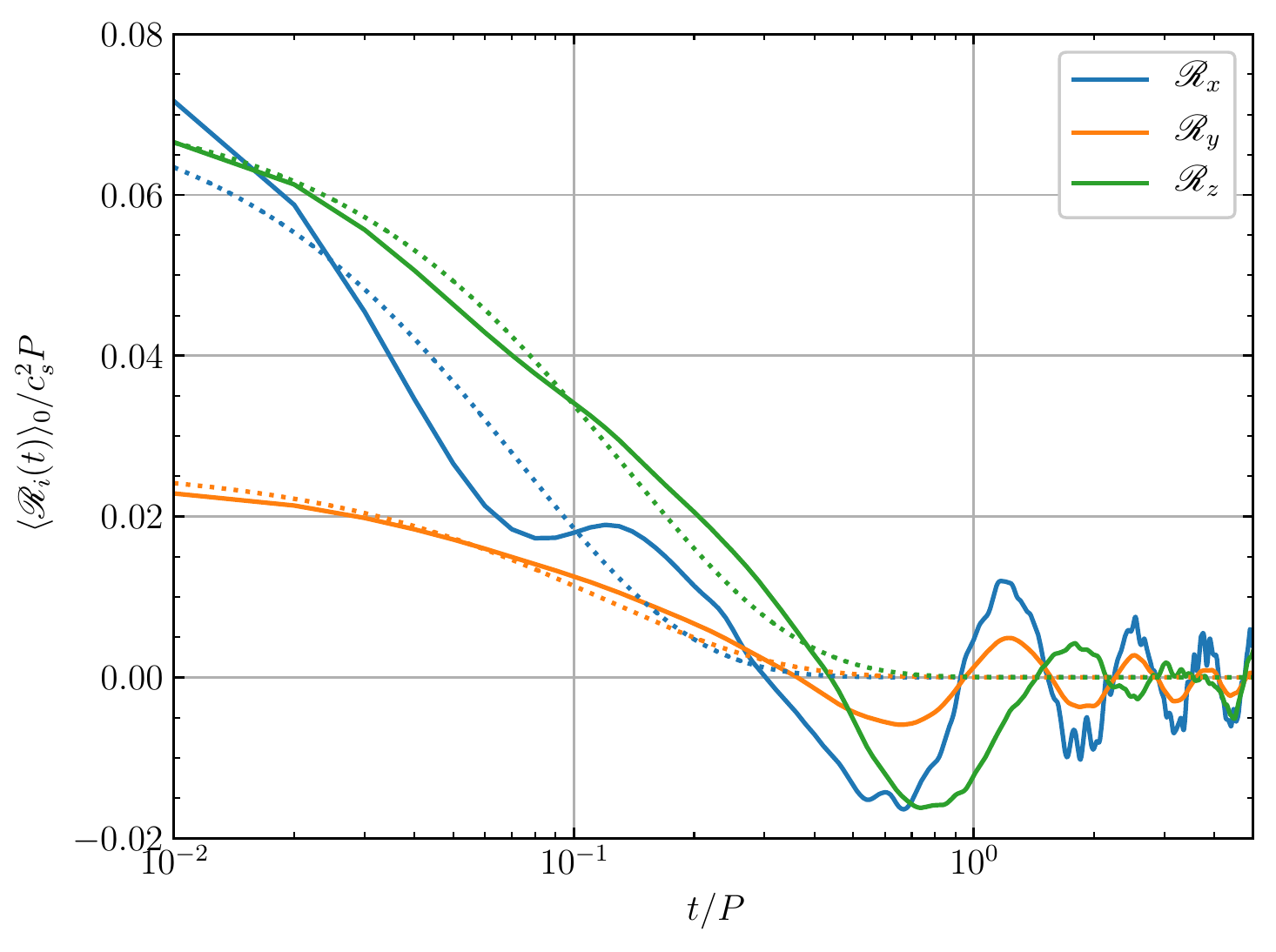}
\caption{Autocorrelation function of the gas velocity fluctuations in the mid-plane of the dead-zone model with a resolution of 32$H_g^{-1}$.
Different color denotes different component of the velocity fluctuations.
The solid lines are horizontal averages computed from the data, while the dotted lines are the best exponential fit to the data.\label{F:uuacf}}
\end{figure}

The solid lines in Figure~\ref{F:uuacf} show the autocorrelation function of the gas velocity fluctuations measured from the mid-plane of the dead-zone model with a resolution of 32$H_g^{-1}$.
The autocorrelation exponentially decays within a time lag of less than a few tenths of an orbital period.
It becomes oscillatory for longer time lags.
The oscillation in the tail of the autocorrelation function indicates that there exist coherent, wavelike motions near the mid-plane of the disk.
The dominant period of these motions appears to be greater than the orbital period $P$.
Hence, the waves passing through the mid-plane may not be purely epicyclic and perhaps consist of several different modes, which is not apparent when considering only the velocity dispersions above.

The autocorrelation for the long time-lag tail prevents us from using integration to estimate the correlation time of the fluctuations, as was done in \cite{YMM09,YMM12} for stochastic torques, because the integration does not lead to satisfactory cancellation over the tail and hence introduces overwhelming numerical errors.
Therefore, we follow the procedure used by \cite{FP06} and fit an exponential function to the measured autocorrelation function.
Shown by the dotted lines in Figure~\ref{F:uuacf}, the fitting is relatively insensitive to the upper limit used for the time lag and gives a more robust estimate of the correlation time from the fitting parameter.
We denote the correlation time in the $i$-th component of the velocity fluctuations at vertical position $z$ by $t_{c,i}(z)$, where $i$ is $x$, $y$, or $z$, and its dimensionless version by $\tau_{c,i}(z) \equiv \OmegaK t_{c,i}(z)$.

\begin{figure}
\plotone{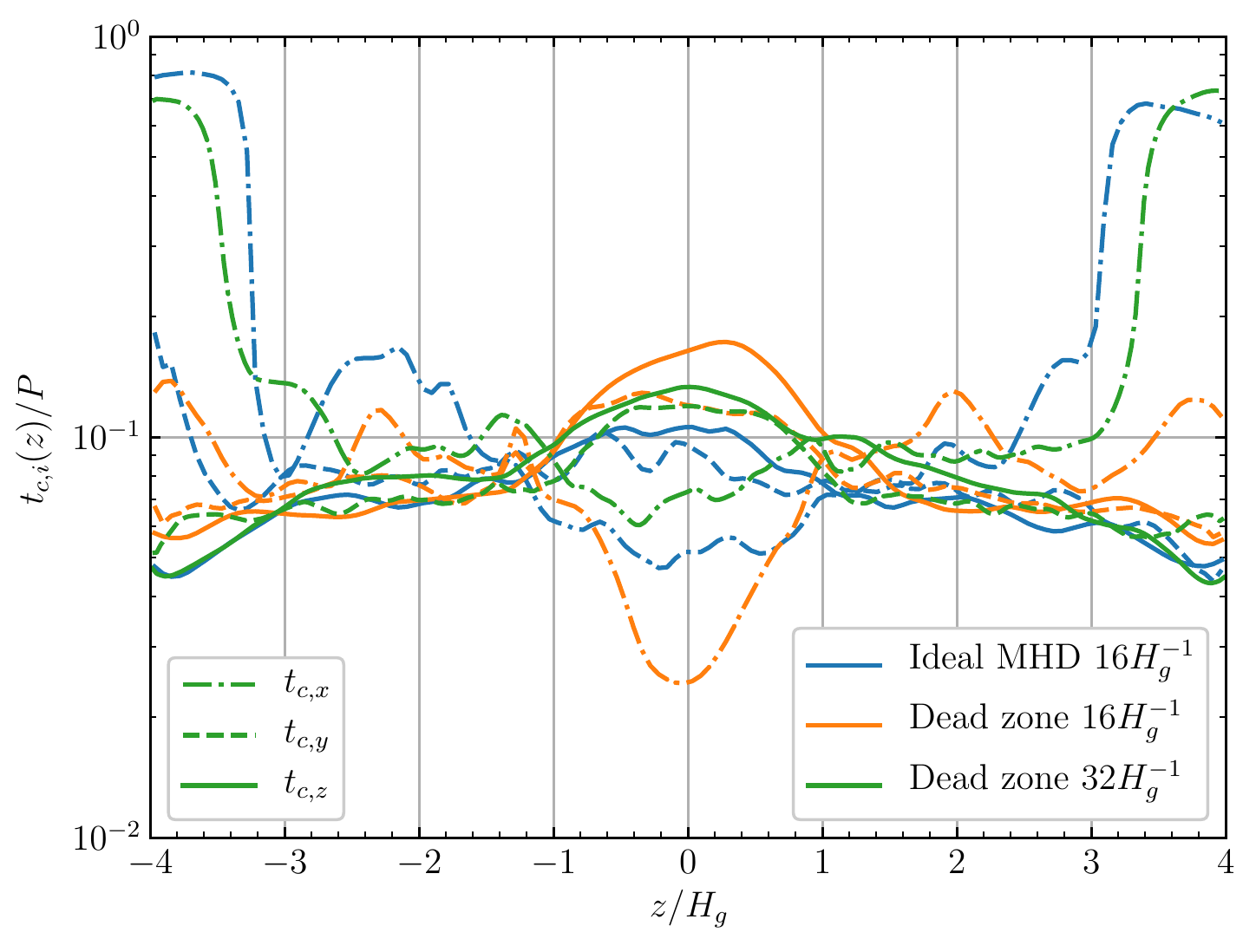}
\caption{Vertical profiles of the estimated correlation time of the gas velocity fluctuations in the saturated state of the flow.
The colors and line styles are the same as in Figure~\ref{F:vpdu}.
The correlation time is normalized by the orbital period $P$.\label{F:vptc}}
\end{figure}

Figure~\ref{F:vptc} shows our estimate of the correlation times $t_{c,i}$ as a function of vertical position in the saturated state of the velocity fluctuations without the back reaction of the solid particles.
In the mid-plane of the ideal-MHD model, $2t_{c,x} \simeq t_{c,y} \simeq t_{c,z} \simeq 0.1P$.
The correlation times in the azimuthal and vertical components are relatively constant over vertical dimension, while that in the radial component significantly increases for $|z| \gtrsim 3H_g$.
Near the mid-plane of the dead-zone models, the correlation times in the azimuthal and vertical components are somewhat longer than their counterparts for the ideal-MHD model, while that in the radial component can be uncertain by a factor of a few.
In the active surface layers, the correlation times in all three components are rather similar, except for the azimuthal component near the vertical boundary $|z| \gtrsim 3H_g$.
The estimated values of the correlation times in the mid-plane of various models are listed in Table~\ref{T:du}.

With both the velocity dispersions $\delta u_i$ and the correlation times $t_{c,i}$ in the velocity fluctuations in hand, we can now estimate the bulk diffusion coefficients $D_{g,i} \sim \overline{\delta u_i}^2 t_{c,i}$ in the saturated state, where $D_{g,i}$ is the diffusion coefficient in the $i$-th direction \citep{FP06,YL07,OH11}.
We can scale the bulk diffusion in each direction following the Shakura--Sunyaev scaling of the shear stress to define the dimensionless bulk diffusion parameters \citep{YL07}:
\begin{equation} \label{E:gasdiff}
\alpha_{g,i}(z) \equiv
\frac{D_{g,i}(z)}{c_s H_g} \simeq
\left[\overline{\frac{\delta u_i(z)}{c_s}}\right]^2 \tau_{c,i}(z),
\end{equation}
where $\alpha_{g,i}(z)$ is a dimensionless measure of the $i$-th diffusion coefficient as a function of vertical position $z$.

\begin{figure}
\plotone{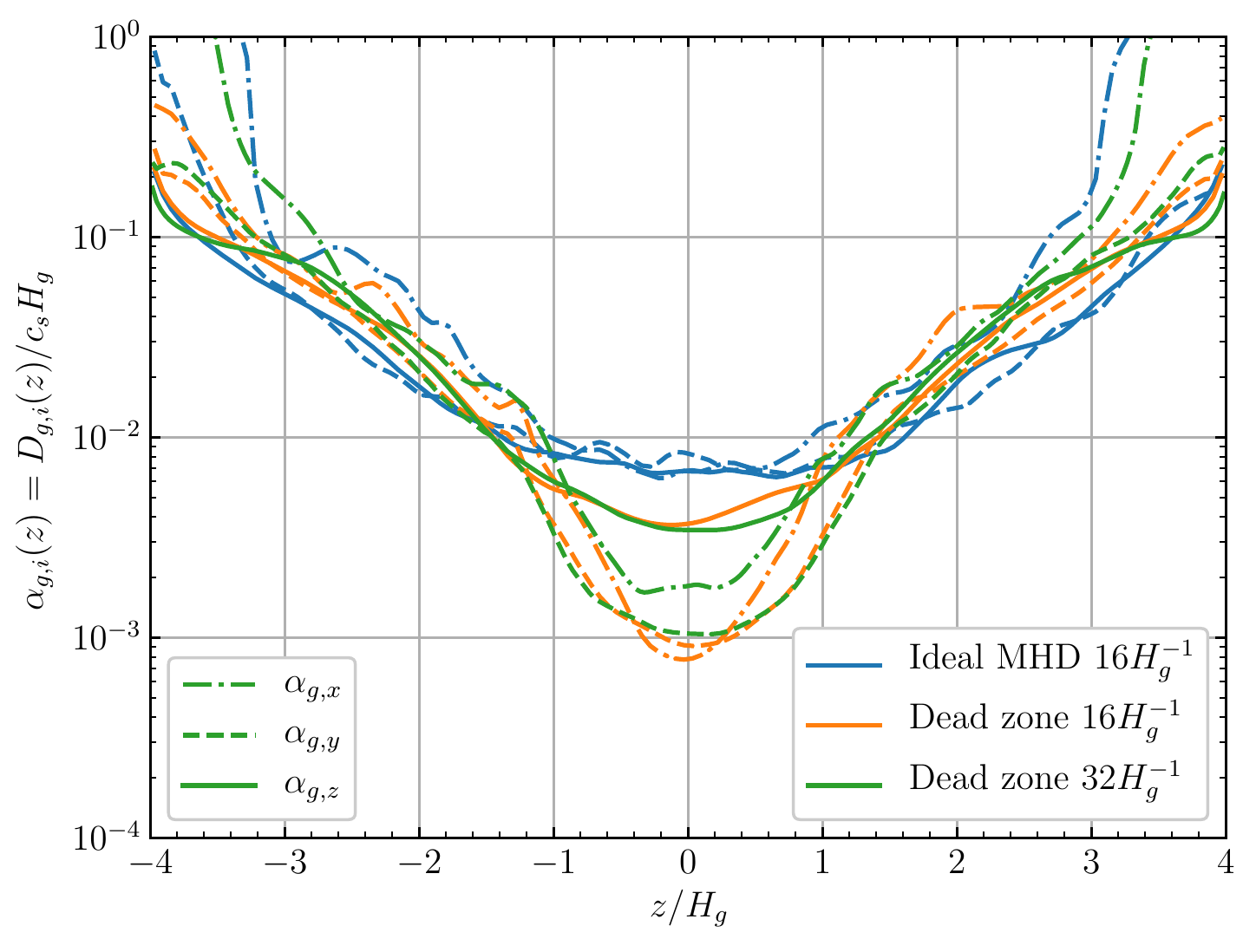}
\caption{Vertical profiles of the estimated diffusion coefficients in the saturated state of the flow, which can be compared with those of the \ssa{} shear stresses in Figure~\ref{F:vpgas}c.
The colors and line styles are the same as in Figure~\ref{F:vpdu}.\label{F:alpha}}
\end{figure}

The resulting vertical profiles of the dimensionless diffusion coefficients $\alpha_{g,i}(z)$ are shown in Figure~\ref{F:alpha}, and the measured values in the mid-plane are listed in Table~\ref{T:du}.
In general, diffusion increases with increasing vertical height in all models (with net vertical magnetic flux).
For the ideal-MHD model, diffusion is rather isotropic up to $|z| \sim 3H_g$ before radial diffusion dominates near the vertical boundary.
In the mid-plane, $\alpha_{g,x}(0) \simeq \alpha_{g,y}(0) \simeq \alpha_{g,z}(0) \simeq \ass(0)$, indicating similar strengths in bulk diffusion and shear stresses.
For the dead-zone model, however, $\ass(0) < \alpha_{g,x}(0) \simeq \alpha_{g,y}(0) < \alpha_{g,z}(0)$.
The bulk diffusion and the shear stresses inside the dead zone, driven by the turbulent surface layers, are not linearly related.
This observation has important consequences in studying the equilibrium vertical distribution of solid particles, as discussed in Section~\ref{SS:parvd}.
Moreover, the appreciably lower radial and azimuthal diffusion in the dead zone may help us understand the clumping of solid particles by the back reaction, as discussed in Sections~\ref{SS:rcd} and~\ref{S:csbr}.
As a final remark, \cite{OH11} suggested that the diffusion coefficients would have a vertical Gaussian profile with a scale length equal to the gas scale height $H_g$.
We note that this may not necessarily be the case, as demonstrated by Figure~\ref{F:alpha}, and the diffusion coefficients may be dependent on the exact resistivity profile.

\section{QUASI-STEADY-STATE PROPERTIES OF THE PARTICLE DISK WITHOUT BACK REACTION}
\label{S:sspar}

We next discuss the properties of the particle disk in the saturated state of the MHD flow without back reaction from the solid particles.
The scale height of the particles is measured and compared with analytical expectation.
Also considered is the concentration and diffusion of solids by the flow in these models, which serves as a baseline to our other models with back reaction presented in Section~\ref{S:csbr}.

\subsection{Vertical Distribution} \label{SS:parvd}

\begin{figure*}
\plotone{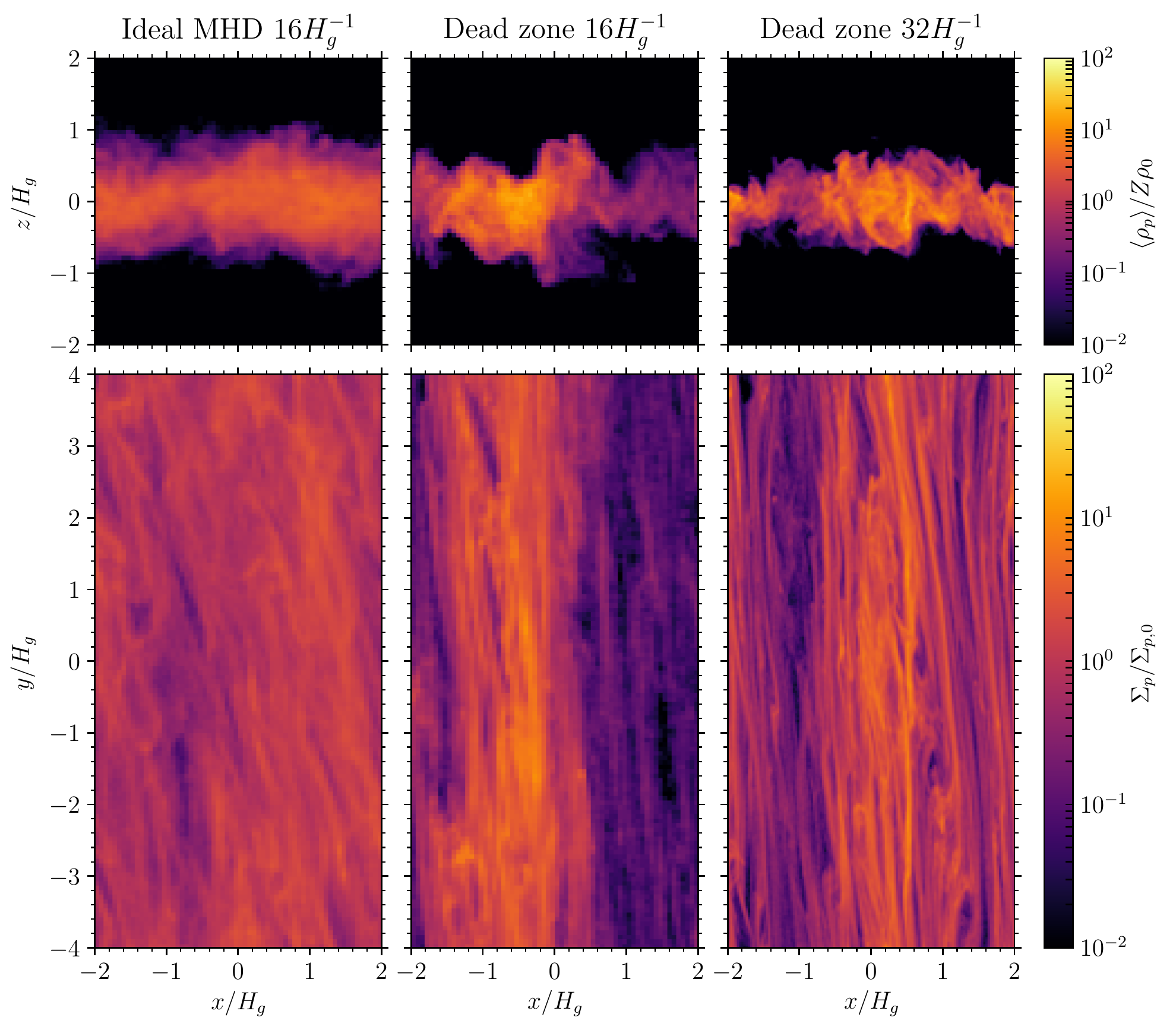}
\caption{Side view (top panels) and top view (bottom panels) of the particle disk at the end of our models without back reaction.
The side view shows the azimuth-averaged particle density $\langle\rho_p\rangle$, while the top view shows the column density of the particles $\Sigma_p$, where $\Sigma_{p,0}$ is the initial column density of the particles.
The region for which $|z| > 2H_g$ is not shown, since no super-particle ever reaches there during the simulation.\label{F:pdnbr}}
\end{figure*}

\begin{figure*}
\plotone{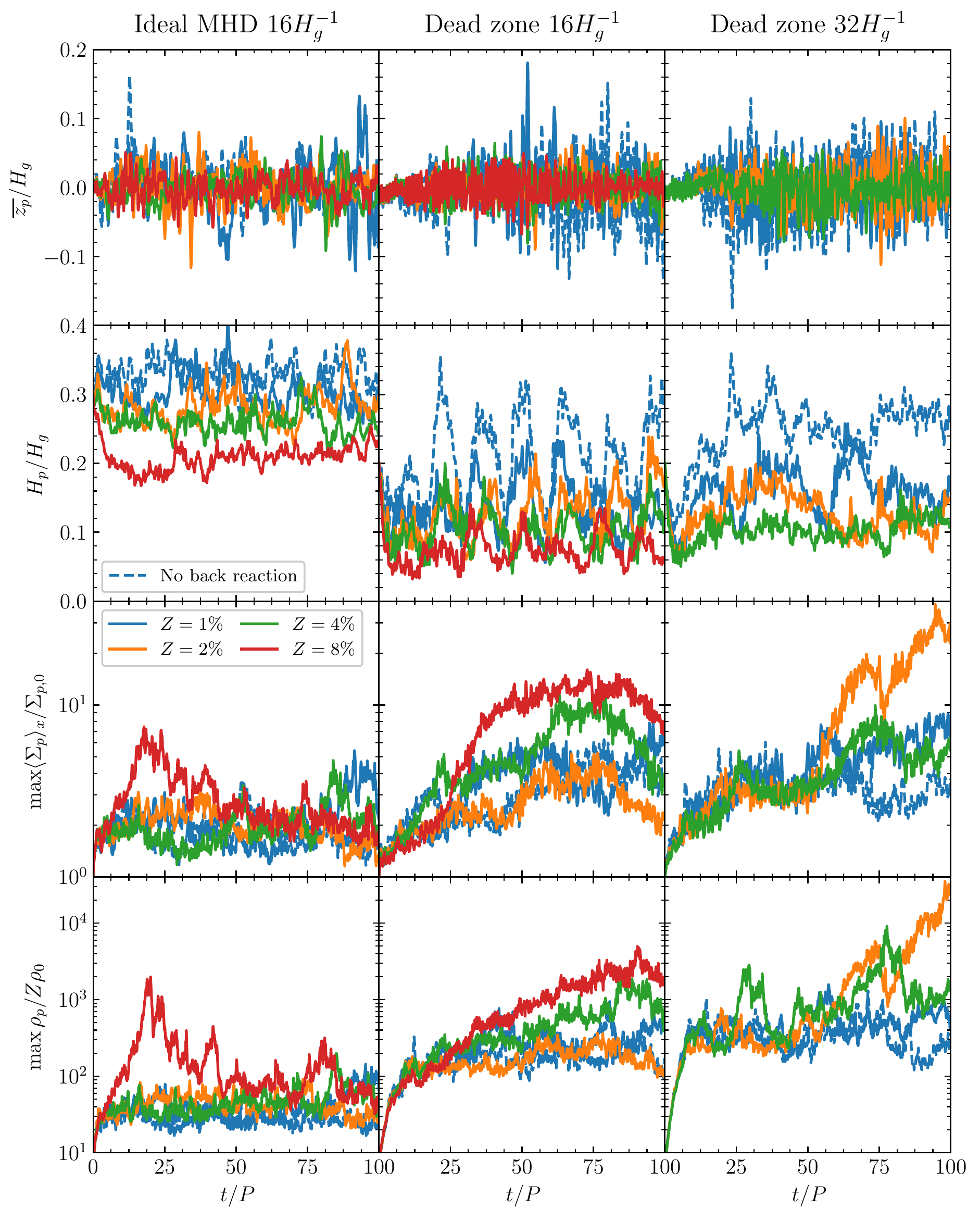}
\caption{Evolution of several diagnostics of the particle distribution for various models.
Each column represents one MHD model at the given resolution.
The panels from top to bottom are the mean vertical center, the vertical scale height, the maximum radial concentration, and the maximum local concentration, respectively.
The dashed line indicates that there is no back reaction of the drag force on the gas, while each solid line denotes a different solid abundance with back reaction in effect.
Noticeable are that the scale height of the particle disk in the dead zone is only a factor of a few less than that in ideal-MHD turbulence, and that the higher the solid abundance, the thinner the disk is.
Moreover, strong local concentration of solid particles is triggered in the dead zone when the solid abundance is a few percent.
\label{F:pdprop}}
\end{figure*}

\begin{deluxetable*}{lCCCCCCCCCCC}
\tabletypesize{\footnotesize}
\tablecaption{Average properties of the particle disk for models without back reaction\label{T:pdnbr}}
\tablehead{
\colhead{Model}
    & \colhead{Resolution}
    & \colhead{$\overline{z_p}$}
    & \colhead{$H_p$}
    & \colhead{Eq.~\eqref{E:hpalpha}\tablenotemark{a}}
    & \colhead{Eq.~\eqref{E:hpduz}\tablenotemark{b}}
    & \colhead{$\max\langle\Sigma_p\rangle_x$}
    & \colhead{$\max\rho_p$}
    & \colhead{$D_{p,x}$}
    & \colhead{$\delta v_{p,x}$}
    & \colhead{$\delta v_{p,y}$}
    & \colhead{$\delta v_{p,z}$}\\
    & \colhead{($H_g^{-1}$)}
    & \colhead{($H_g$)}
    & \colhead{($H_g$)}
    &
    &
    & \colhead{($\Sigma_{p,0}$)}
    & \colhead{($10^2 Z\rho_0$)}
    & \colhead{($c_s H_g$)}
    & \colhead{($c_s$)}
    & \colhead{($c_s$)}
    & \colhead{($c_s$)}
}

\startdata
Ideal MHD & 16 & +0.00(3) & 0.32(2) & 0.28 & 0.26
    & 1.8(4)       & 0.29(8)          & 1.7\times10^{-2}
    & 0.14(1) & 0.10(1)\phn & 0.090(9)\\
Dead Zone & 16 & -0.01(5) & 0.20(7) & 0.05 & 0.19
    & 4(1)\phd\phn & 2(1)\phd\phn\phn & 7.0\times10^{-4}
    & 0.06(2) & 0.03(1)\phn & 0.05(2)\phn\\
Dead Zone & 32 & +0.00(4) & 0.25(3) & 0.05 & 0.18
    & 3.2(6)       & 3(2)\phd\phn\phn & 1.2\times10^{-3}
    & 0.06(1) & 0.038(6)    & 0.06(1)\phn\\
\enddata

\tablecomments{The time average is taken from $t_1 = 40P$ to $t_2 = 100P$.
    The one standard deviation over time for each property is shown in parentheses.}
\tablenotetext{a}{Analytical estimate of $H_p / H_g$ using \ssa{} stress parameter $\ass$.}
\tablenotetext{b}{Analytical estimate of $H_p / H_g$ using vertical velocity fluctuations of the gas.}

\end{deluxetable*}

The top panels of Figure~\ref{F:pdnbr} show the side view of the particle disk at the end of the simulation ($t = 100P$) for our models without back reaction.
Although there is radial and vertical substructure in the distribution of particles, the vertical distribution when horizontally averaged is well approximated by a Gaussian function.
The dashed lines in the first row of Figure~\ref{F:pdprop} show the evolution of the vertical center of the particles in the respective models.
The vertical center of the particles is not stationary, but undergoes oscillations with the local Keplerian frequency.
Even inside the dead zone, the velocity fluctuations can lift the center of the particle disk to $\gtrsim$10\% of the gas scale height $H_g$, a length scale which is resolved in our models.

The dashed lines in the second row of Figure~\ref{F:pdprop} show the evolution of the scale height of the particle disk measured in our models without back reaction.
The timescale for the disk to reach equilibrium scale height is governed by $P / (2\pi\tau_s)$ for $\tau_s \ll 1$ \citep[e.g.,][]{DMS95,JK05}, which is $\sim$2$P$ in our case.
For the ideal-MHD model, the scale height of the particles remains fairly steady at $\sim$0.3$H_g$ with a relatively small variation of amplitude $\sim$0.03$H_g$.
On the other hand, the scale height of the particles for our dead-zone models is on the level of $\sim$0.2$H_g$ and has a stronger variation of amplitude $\sim$0.04--0.1$H_g$ on a longer timescale.
Our measured mean center and scale height of the particle disk is listed in Table~\ref{T:pdnbr} along with their standard deviation over time.

For comparison, the layer of particles in numerical simulations of ambipolar diffusion regulated flow seems to be thinner compared with what we find in an Ohmic dead zone.
In simulations with a net vertical magnetic flux of $\beta_0 \simeq 10^4$ and an ambipolar diffusion number of $\mathrm{Am} \simeq 1$ in the mid-plane, where $\mathrm{Am}$ is the number of times a neutral particle collides with ions during $\OmegaK^{-1}$ \citep{HS98,CM07}, the measured scale height of the particles of $\tau_s = 0.1$ covers a range of values from $\sim$0.04 to $\sim$0.1$H_g$ \citep{ZSB15,XBM17,RL18}.
This is smaller than our measured value of $\sim$0.2$H_g$, but remains noticeably larger than what streaming turbulence alone supports at $\sim$0.02$H_g$ \citep{CJD15}.

We now evaluate some analytical estimates of the scale height of the particle disk from the properties of the MHD flow and compare them with our measured values.
First, we consider the estimate using the \ssa{} stress parameter, i.e., turbulent shear stresses \citep{DMS95}:
\begin{equation} \label{E:hpalpha}
\frac{H_p}{H_g} \simeq \sqrt{\frac{\ass(0)}{\tau_s + \ass(0)}}.
\end{equation}
We use the measured $\ass$ values at the mid-plane in Figure~\ref{F:vpgas}c and Table~\ref{T:du}, and the results are listed in the fifth column of Table~\ref{T:pdnbr}.
This estimate yields $\sim$0.3 for the ideal-MHD model, in good agreement with the measured one.
However, the estimate for the dead-zone models is only 0.05, four or five times lower than the measured ones.
These low estimates are due to the low stresses inside the dead zone.
It is difficult to attribute this discrepancy to the uncertainty in the leading coefficient in Equation~\eqref{E:hpalpha} since the ideal-MHD model renders a relatively accurate estimate.

\cite{YL07} have cautioned that the $\alpha$ parameter in Equation~\eqref{E:hpalpha} should not be interpreted as the turbulent shear stresses, as assumed by \cite{DMS95}, but rather as the vertical bulk diffusion in the gas due to the vertical velocity fluctuations.
We therefore use the coefficients of vertical bulk diffusion measured in Figure~\ref{F:alpha} and Table~\ref{T:du} to estimate the scale height of the particles instead \citep[see also][]{CFP06,FP06,OH11}:
\begin{equation} \label{E:hpduz}
\frac{H_p}{H_g} \simeq
\sqrt{\frac{\alpha_{g,z}(0)}{\tau_s + \alpha_{g,z}(0)}}.
\end{equation}
The results are listed in the sixth column of Table~\ref{T:pdnbr}.
The ideal-MHD model gives a value of 0.26, only slightly lower than the measured $0.33\pm0.02$.
On the other hand, the dead-zone models give rather accurate estimates of 0.19 and 0.18 for the resolutions of 16$H_g^{-1}$ and 32$H_g^{-1}$, in comparison with the measured $0.20\pm0.07$ and $0.25\pm0.03$, respectively.
This exercise strengthens the dichotomy between the bulk diffusion and shear stresses, especially when considering particle-gas dynamics inside the dead zone, emphasizing that Equation~\eqref{E:hpalpha} should not be used, but rather Equation~\eqref{E:hpduz}.
We note that \cite{ZSB15} and \cite{XBM17} found similar results for the case of MHD turbulence driven by ambipolar diffusion.

Finally, even though horizontally averaged vertical distribution of particles in MHD turbulence can be well understood, substructures do exist across both the radial and vertical dimensions (top panels of Figure~\ref{F:pdnbr}).
For the ideal-MHD model, the particles are relatively well mixed, with variations only on longer spatial scales.
On the other hand, the dead-zone models demonstrate apparently localized structures, and this feature further enhances with increasing resolution.
This emphasizes that the mid-plane flow is no longer a uniform turbulent flow.
Nevertheless, because the waves excited by the turbulent surface layers before they propagating into the mid-plane are spatially local \citep{BS13} and temporally random, the perturbations of the gas inside the dead zone still constitute a random process on average and hence drive the diffusion of the solid particles.

\subsection{Radial Concentration and Diffusion} \label{SS:rcd}

With the quasi-steady vertical distribution of particles discussed in Section~\ref{SS:parvd}, we next turn to their concentration and diffusion in the radial direction.
This can be illustrated by the bottom panels of Figure~\ref{F:pdnbr}, which shows the top view of the particle disk at the end of each model without back reaction.
Due to the background shear flow, the spatial variations are predominantly radial, and this is especially apparent in the dead-zone models, where perturbations in the gas excited from the active surface layer experience even more shear when propagating down into the mid-plane \citep{OO13}.

\begin{figure*}
\plotone{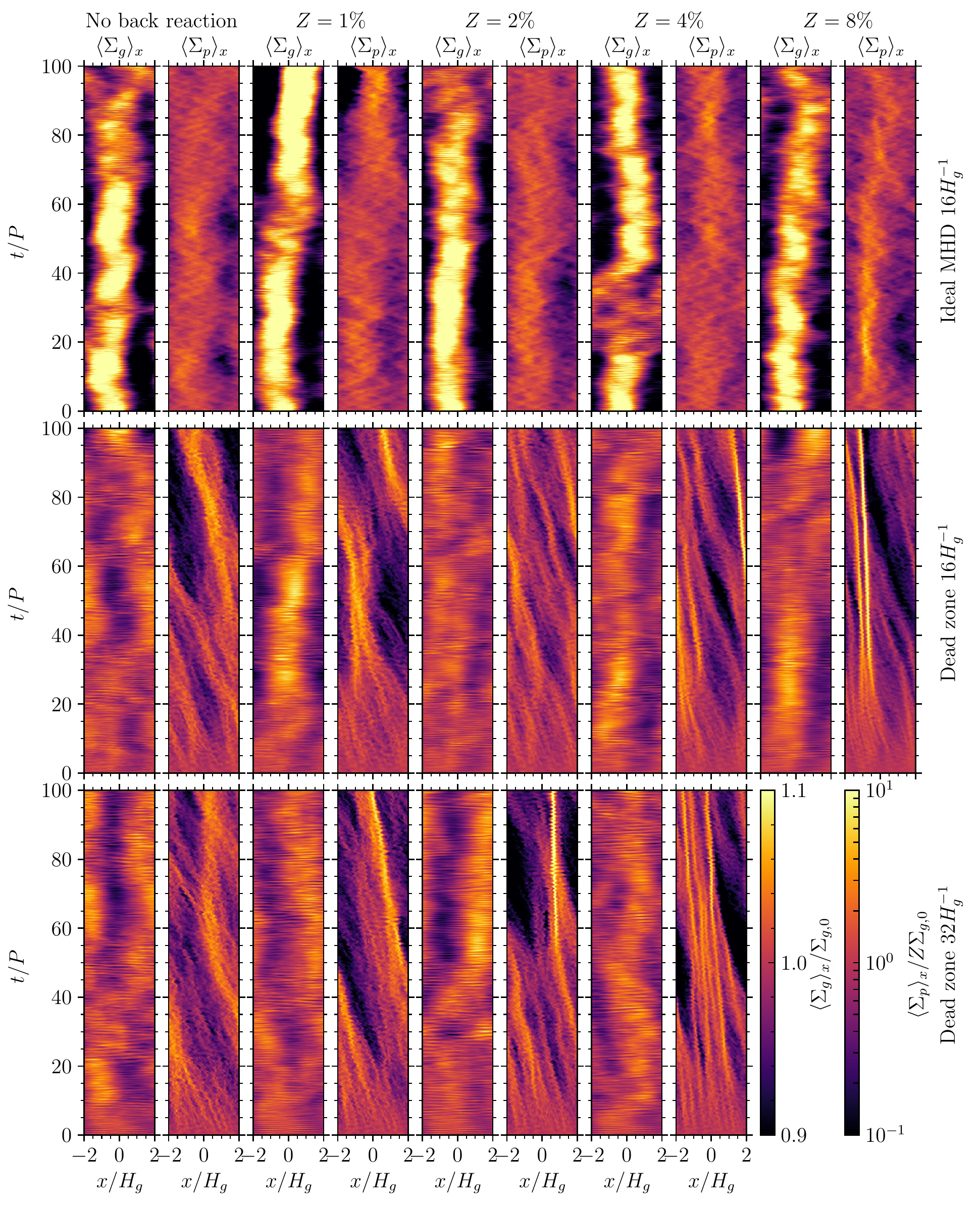}
\caption{Azimuthally averaged column densities of the gas and the particles as a function of radial position $x$ and time $t$ for various models.
Each panel consists of two images with the gas and the particle column densities on the left and the right, respectively, and their magnitude are indicated by the colorbars on the lower right corner.
Each row represents one MHD model at the given resolution, while each column denotes different solid abundance.
\label{F:sigmaxt}}
\end{figure*}

The two left-most columns in Figure~\ref{F:sigmaxt} show the azimuthally averaged column densities of the gas $\langle\Sigma_g\rangle_x$ and of the particles $\langle\Sigma_p\rangle_x$ as a function of radial position $x$ and time $t$ for each model without back reaction.
For the ideal-MHD model, the mode of the first harmonics (i.e., the longest wavelength a finite dimension can represent) dominates the perturbations in the gas, with an amplitude of $\sim$12\% \citep[cf.,][]{JYK09,YMM12}.
Meanwhile, the distribution of the solid particles is well correlated with that of the gas, driving a radial concentration of about a factor of two.
For the dead-zone models, the mode of the first harmonics also dominates, however with an appreciably smaller amplitude of $\sim$4\%.
By contrast, the concentration of the solids is slightly stronger, by about a factor of three.
More importantly, the solids concentrate into relatively narrow bands (see also Figure~\ref{F:pdnbr}), and are less well correlated with the gas than in the ideal-MHD case.
The maximum $\langle\Sigma_p\rangle_x$ over radial position as a function of time for each model is plotted as the dashed line in the third row of Figure~\ref{F:pdprop}, and the corresponding time average is listed in the seventh column of Table~\ref{T:pdnbr}.

The dashed lines in the bottom row of Figure~\ref{F:pdprop} show the maximum local density of solids as a function of time for our models without back reaction.
By comparing with the maximum azimuthal average of column density in the third row, the maximum local density for the ideal-MHD model correlates well with the radial concentration, while the correlation is poorer in the dead-zone models.
Moreover, the level of the maximum local density reached cannot be accounted for by the combination of radial concentration and vertical sedimentation only, indicating the presence of some level of azimuthal concentration, which can also be seen in Figure~\ref{F:pdnbr}.
As listed in the eighth column of Table~\ref{T:pdnbr}, the local concentration of solids without back reaction for the dead-zone models is an order of magnitude stronger than that for the ideal-MHD model.

We further measure the coefficient of radial diffusion of solid particles $D_{p,x}$ in these models by following the procedure used in \cite{YMM09}.
We record the radial displacement of each particle and compute its distribution as a function of time, which resembles a Gaussian function.
The diffusion coefficient can then be estimated by fitting a $\sqrt{t}$ function to the width of the distribution.
The results are listed in the ninth column of Table~\ref{T:pdnbr}.
Interestingly, the radial diffusion of particles in the dead-zone models is more than an order of magnitude weaker than in the ideal-MHD model.
This is consistent with the significantly finer radial variations in the column density of particles, as shown in Figure~\ref{F:pdnbr}, and the significantly lower radial diffusion in the gas, as shown in Table~\ref{T:du}.

Finally, listed in Table~\ref{T:pdnbr} are the components of the velocity dispersion of the particles $\delta\vec{v}_p$ measured from each model.
The measured values are close to those for the gas listed in Table~\ref{T:du}, which is expected from the relatively tight coupling between the gas and the particles \citep{YL07}.
The magnitude of the velocity dispersion for the ideal-MHD model is about 0.19 the speed of sound $c_s$, while that for the dead-zone models is about 0.09$c_s$.
We note that this magnitude is comparable to or more than the difference between the gas velocity and the Keplerian velocity $\Delta u_y = \Pi c_s = 0.05c_s$ driven by the background radial pressure gradient (Sections~\ref{SSS:MHD} and~\ref{SS:ibc}).

\section{CONCENTRATION OF SOLIDS DRIVEN BY BACK REACTION} \label{S:csbr}

In the preceding section, we focus on the particle-gas dynamics in MHD turbulence where solid particles do not exert drag force on the gas, and hence the particles are only passively pushed around by the flow.
This analysis provides a baseline for how strongly solid materials can sediment and be concentrated by the fluctuating gas.
In this section, we activate the back reaction of the drag force from the particles on the gas and study its effects together with flow-driven diffusion and concentration.
We systematically increase the solid abundance from $Z = 0.01$ up to $Z = 0.08$, which is equivalent to increasing the importance of the back reaction.

\begin{deluxetable*}{lCCCCCCCC}
\tablecaption{Average properties of the particle disk for models with back reaction\label{T:pdbr}}
\tablehead{
\colhead{Model}
    & \colhead{Resolution}
    & \colhead{$Z$}
    & \colhead{$\overline{z_p}$}
    & \colhead{$H_p$}
    & \multicolumn{2}{c}{$\max\langle\Sigma_p\rangle_x$}
    & \multicolumn{2}{c}{$\max\rho_p$}\\
    & \colhead{($H_g^{-1}$)}
    &
    & \colhead{($H_g$)}
    & \colhead{($H_g$)}
    & \multicolumn{2}{c}{($\Sigma_{p,0}$)}
    & \multicolumn{2}{c}{($10^2 Z\rho_0$)}\\
    & &
    & \colhead{Average\tablenotemark{a}}
    & \colhead{Average\tablenotemark{a}}
    & \colhead{Average\tablenotemark{a}} & \colhead{Maximum\tablenotemark{b}}
    & \colhead{Average\tablenotemark{a}} & \colhead{Maximum\tablenotemark{b}}
}

\startdata
Ideal MHD & 16
    & 0.01 & -0.00(5) & 0.30(3)
           & \phn2.1(9)       & \phn5.5
           & \phn0.4(2)       & \phn\phn1.4\\
  & & 0.02 & -0.00(2) & 0.29(3)
           & \phn1.8(4)       & \phn3.0
           & \phn0.4(1)       & \phn\phn1.1\\
  & & 0.04 & -0.00(2) & 0.26(2)
           & \phn2.1(6)       & \phn4.8
           & \phn0.5(2)       & \phn\phn2.0\\
  & & 0.08 & -0.00(2) & 0.21(1)
           & \phn2.2(5)       & \phn4.0
           & \phn0.8(5)       & \phn\phn4.9\\
Dead zone & 16
    & 0.01 & +0.00(4) & 0.13(5)
           & \phn4(1)\phd\phn & \phn8\phd\phn
           & \phn3(1)\phd\phn & \phn\phn8\phd\phn\\
  & & 0.02 & +0.00(2) & 0.13(4)
           & \phn3.1(9)       & \phn5.2
           & \phn1.9(7)       & \phn\phn3.7\\
  & & 0.04 & -0.00(2) & 0.10(3)
           & \phn6(3)\phd\phn &    12\phd\phn
           & \phn6(4)\phd\phn &    \phn22\phd\phn\\
  & & 0.08 & +0.00(2) & 0.07(2)
           &    11(2)\phd\phn &    16\phd\phn
           &    14(10)\phd    &    \phn53\phd\phn\\
Dead zone & 32
    & 0.01 & -0.00(4) & 0.16(3)
           & \phn5(2)\phd\phn & \phn9\phd\phn
           & \phn5(3)\phd\phn &    \phn16\phd\phn\\
  & & 0.02 & -0.00(3) & 0.12(3)
           &    10(13)\phd    &    39\phd\phn
           &    19(52)\phd    &       371\phd\phn\\
  & & 0.04 & -0.00(3) & 0.10(2)
           & \phn5(2)\phd\phn &    10\phd\phn
           &    11(11)\phd    &    \phn91\phd\phn\\
\enddata

\tablecomments{The standard deviation over time for each property is shown in parentheses.}
\tablenotetext{a}{Time average from $t_1 = 40P$ to $t_2 = 100P$.}
\tablenotetext{b}{Absolute maximum from $t_1 = 40P$ to $t_2 = 100P$.}

\end{deluxetable*}

The solid lines in the second row of Figure~\ref{F:pdprop} show the scale height of the particle disk as a function of time for various MHD models and solid abundances.
When $Z = 0.01$, the particles in the ideal-MHD model have a similar level and similar variations in scale height as the case without back reaction.
For the dead-zone models, on the other hand, the case of $Z = 0.01$ demonstrates noticeable further sedimentation compared to the case without back reaction.
In any case, both the level and the variations in scale height of the particle disk decrease with increasing solid abundance.
This behavior is consistent with previous simulations without MHD turbulence \citep{CJD15,YJC17}.
Physically, the dependence of the particle scale height on solid abundance may be understood because the combination of mutual drag force and the solid loading (in the limit of small stopping time) effectively reduces the speed of sound in the dust-gas mixture \citep{SC13,LY17}.
In addition, the vertical center of the particle disk undergoes vertical oscillations as in the case without the back reaction, as shown by the solid lines in the first row of Figure~\ref{F:pdprop}.
The amplitude of the oscillations also decreases with increasing solid abundance, which is a natural consequence of the reducing scale height of the particles.
The time average and variation of the vertical center and scale height of the particle disk for various models is listed in Table~\ref{T:pdbr}.

We are finally in a position to address the main question of this work:
Can the solid particles of dimensionless stopping time $\tau_s = 0.1$ spontaneously concentrate via the mutual drag force in MHD turbulence?
To quantify this, we scale the column and local densities of the solids with the solid abundance $Z$ in Figures~\ref{F:pdprop} and~\ref{F:sigmaxt} so that the strength of self-induced concentration for different abundances can be compared with each other as well as to the case without the back reaction.
From the scaled densities, we compute in Table~\ref{T:pdbr} the time average and the absolute maximum over the duration from $t_1 = 40P$ to $t_2 = 100P$ of the maximum azimuthally-averaged column density $\max\langle\Sigma_p\rangle_x$ and the maximum local density $\max\rho_p$ of the solids.
The former reveals the strength of radial concentration while the latter indicates the local concentration combined in all three dimensions.

For the ideal-MHD model, it appears that the back reaction does not enhance the concentration of such particles for solid abundance below $Z \sim 0.04$.
The level of radial and local concentration is rather similar as in the case without the back reaction.
Moreover, it seems that the radial concentration of the solids also correlates well with that of the gas, in which the mode of the first harmonics dominates (see Section~\ref{SS:rcd}), as shown in the first row of Figure~\ref{F:sigmaxt}.
For the solid abundance of $Z = 0.08$, a transient, strong concentration of solids does appear around $t \sim 20P$, forming one dense axisymmetric filament, but is dispersed soon afterwards.
The local concentration of solids then stays at a slightly higher level than the cases with $Z \lesssim 0.04$, without formation of any major filament of solids.

For the dead-zone model, we first consider the same resolution of 16$H_g^{-1}$ as used in the ideal-MHD model (the second column of Figure~\ref{F:pdprop} and the second row of Figure~\ref{F:sigmaxt}).
When the solid abundance $Z = 0.01$ or $Z = 0.02$, the level of radial and local concentration remains similar to that in the case without the back reaction.
In all three cases, one or two relatively broad, loose, and axisymmetric filaments of solids can be seen in the evolution.
Some further clumping of solids appears intermittently when back reaction is in effect.
On the other hand, the cases of $Z = 0.04$ and $Z = 0.08$ begin to show appreciable further concentration of solids driven by the back reaction.
One or two dominant axisymmetric filaments emerge and maintain their dominance to the end of the simulations.
The strength of the concentration scales roughly linearly with the solid abundance with respect to the case of $Z = 0.02$ (see Table~\ref{T:pdbr}); in combination, this results in a $Z^2$ increase in the absolute density of solids.

We proceed to consider the higher resolution of 32$H_g^{-1}$ for the dead-zone model up to a solid abundance of $Z = 0.04$.
The quantitative dependence of solid concentration on $Z$ is less clear.
However, transient but significantly larger local concentrations of solids do appear.
For the cases of $Z = 0.01$ and $Z = 0.04$, the average strength of radial concentration of solids is about a factor of five, which is slightly stronger than the case without the back reaction (Figure~\ref{F:pdprop} and Table~\ref{T:pdbr}).
One and three narrow axisymmetric filaments of solids exist in the respective cases most of the time in the simulations, which are absent in the case without back reaction (Figure~\ref{F:sigmaxt}).
In addition, the local concentration in the case of $Z = 0.04$ is about a factor of two stronger than that in the case of $Z = 0.01$ on average.

\begin{figure*}
\plotone{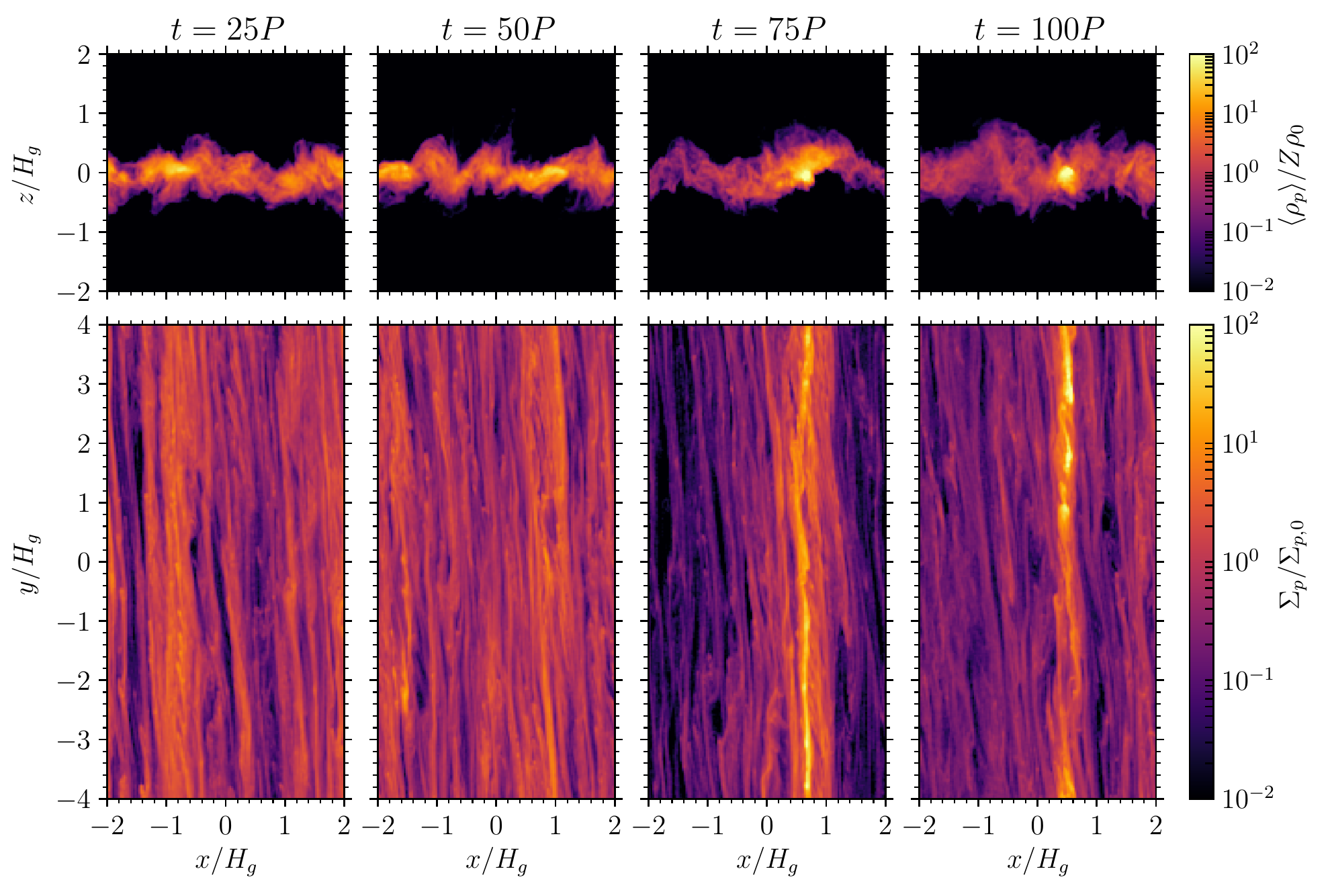}
\caption{Evolution of the particle disk in our dead-zone model with a resolution of 32$H_g^{-1}$ and a solid abundance of $Z = 0.02$.
The top panels show the side view of the disk, while the bottom panels show the top view; the time increases from left to right.
The color scales are the same as those used in Figure~\ref{F:pdnbr} for comparison purposes.\label{F:pdevol}}
\end{figure*}

The dead-zone model with a resolution of 32$H_g^{-1}$ and a solid abundance of $Z = 0.02$ presents a particularly interesting case.
Figure~\ref{F:pdevol} shows the evolution of the particle disk in this case.
A strong axisymmetric filament of solids forms at $t \sim 55P$ and continues to accumulate more solids afterwards.
The strength of radial concentration reaches a factor of about 40, while the strength of local concentration reaches about $4\times10^4$ (Figure~\ref{F:pdprop} and Table~\ref{T:pdbr}).
This level of solid concentration is much larger than the cases of $Z = 0.01$ and $Z = 0.04$.

Even though it remains difficult to exactly quantify the solid concentration driven by the back reaction, it seems clear that solid loading does enhance the spontaneous concentration of solid particles to high density in the dead zone.
For $Z \gtrsim 0.08$ at a resolution of 16$H_g^{-1}$ and $Z \gtrsim 0.02$ at a resolution of 32$H_g^{-1}$, the peak local solid density reaches more than 200$\rho_0$, where $\rho_0$ is the background gas density in the mid-plane.
Even for the case of $Z = 0.04$ at a resolution of 16$H_g^{-1}$, a peak local solid density of $\sim$90$\rho_0$ is reached.
These densities are well over the Roche density in most parts of a typical protoplanetary disk \citep{YJC17} and thus the formation of planetesimals via gravitational collapse should proceed.
We note that the resolutions we have considered in this work are not sufficient to resolve the critical wavelength of the linear streaming instability, and hence the dynamical timescale of the system should become shorter in models with higher resolution.

The actual critical solid abundance for clumping and ultimately planetesimal formation appears to depend on both the disk magnetization and the stopping time.
Large particles with $\tau_s = 1$ were found by \citet{JO07} to clump with $Z \gtrsim 0.01$ in the ideal MHD case, but only with $Z \gtrsim 0.03$ for the hydrodynamic case.
In the models reported here with $\tau_s = 0.1$, on the other hand, we find that the ideal MHD case only produces clumping with $Z \gtrsim 0.08$, but the pure hydrodynamic case was found by \citet{CJD15} to clump already with $Z \gtrsim 0.015$.
Thus it remains unclear whether increasing resolution would result in a markedly different critical abundance.
In any case, a solid abundance of a few percent in the dead zone seems sufficient to drive the formation of planetesimals.

We emphasize that strong clumping of solids only occurs in the nonlinear phase of the streaming instability.
Linear modes of the instability do not describe any traffic jam 
\citep{YJ07,JBL11}, but only act as a source of energy to drive random motion and diffusion in the solid particles \citep{JY07}.
The nonlinear phenomenon of the traffic jam driven by the mutual drag force can be intuitively understood by secular accumulation of solids onto increasingly slower drifting clumps \citep[see, e.g.,][Section~1]{YJ14}.
A more formal description of this phenomenon should be developed in future work.

Finally, we note that the tendency for solid particles to concentrate may be a direct consequence of the strength of the radial diffusion in the MHD turbulence.
As discussed in Sections~\ref{SS:vf} and~\ref{SS:rcd}, the vertical diffusion of the gas and the particles near the mid-plane is relatively similar between the ideal-MHD and the dead-zone models, but the radial diffusion is much weaker in the dead-zone models.
As shown in this section, the ideal-MHD model shows no significant concentration of solids for $Z \lesssim 0.04$, unlike the dead-zone model at the same resolution.
On the other hand, for the case of $Z = 0.08$ in the ideal-MHD model, some transient strong local concentration does appear.
Given that higher resolutions do tend to enhance the solid concentration and lower the critical threshold of solid abundance for concentration \citep{YJ14,YJC17}, as also seen in the dead-zone models in this work, it remains possible that solid particles of dimensionless stopping time $\tau_s = 0.1$ may spontaneously concentrate themselves in the ideal MHD turbulence at a moderately higher solid abundance than in the dead zone.

\section{CONCLUDING REMARKS} \label{S:cr}

In this work, we use local-shearing-box simulations to study the streaming instability in a dead zone.
We model a particle-gas system with mutual drag interactions between the gas and the solid particles, including MHD turbulence in a protoplanetary disk.
We systematically compare models with ideal MHD and inside an Ohmic dead zone, with and without back reaction of the solid particles to the gas drag, and with varying solid abundance.
We find that the turbulence in gas near the mid-plane of disks with ideal MHD is relatively isotropic, and the strength of the resulting turbulent diffusion is comparable to that of the accretion stress, i.e., $\ass(0) \simeq \alpha_{g,x}(0) \simeq \alpha_{g,y}(0) \simeq \alpha_{g,z}(0)$, where $\ass(z)$ is the \cite{SS73} viscous accretion stress parameter as a function of vertical position $z$, and $\alpha_{g,i}(z)$ is the dimensionless turbulent diffusion coefficient in the $i$-th direction as a function of $z$.
On the other hand, the velocity fluctuations in the gas inside the dead zone are noticeably anisotropic, and there is a significant dichotomy between diffusion and accretion stress, i.e., $\ass(0) < \alpha_{g,x}(0) \simeq \alpha_{g,y}(0) < \alpha_{g,z}(0)$, where the accretion stress is about an order of magnitude weaker than the vertical diffusion.
Moreover, the strength of vertical diffusion in the model with ideal MHD and inside the dead zone only differs by order unity.
This results in rather similar equilibrium scale heights of the particle disks for the two cases.
Therefore, caution needs to be exercised when considering the relationship between diffusion and accretion stress in protoplanetary disks with non-ideal MHD and using it to estimate the vertical scale height of the solid particles.

Even though solid particles of dimensionless stopping time $\tau_s = 0.1$ do not sediment into the mid-plane of a dead zone appreciably more than their counterparts in fully developed magneto-rotational turbulence, the back reaction of the solid particles to the gas drag remains effective in driving strong radial concentration of the solids inside the dead zone.
A solid abundance of $Z \sim 2\%$ allows these un-sedimented solids to spontaneously concentrate to densities that are over the Roche density, sufficient to lead to the formation of planetesimals.
The relative ease of triggering strong concentration of these solids in a resistive dead zone compared to ideal magneto-rotational turbulence can be explained by the appreciably weaker diffusion of particles in the radial compared to the vertical direction inside the dead zone.
Although the quantitative threshold may be resolution dependent, the qualitative result appears robust.

We remark that even if the initial solid abundance of a disk is below the critical condition, it can still be enhanced by photoevaporation of the outer disk \citep{CG17,EJ17}, ice sublimation and condensation near the ice line \citep{RJ13,IG16,DA17,SO17}, or radial pile-up of solids \citep{DAM16,GLM17}.
We note also that in general, the higher the solid abundance, the more effective sedimentation of the solid particles is, and larger number of dense axisymmetric filaments of solids are formed, which is consistent with models without external turbulence \citep{YJC17}.

This work indicates that the effectiveness of the back reaction to drive strong concentration of solids is not sensitive to the vertical sedimentation of the solid particles.
The scale height of particles of $\tau_s = 0.1$ under MHD turbulence is $\sim$0.2--0.3\,$H_g$, where $H_g$ is the vertical scale height of the gas, while that of similar particles without external turbulence is $\sim$10$^{-2}H_g$ \citep{YJ14,CJD15}, more than an order of magnitude thinner.
However, particles both inside a dead zone and without external turbulence similarly require a solid abundance of a few percent to trigger strong concentration \citep{CJD15,YJC17}.
This implies that the local condition $\rho_p \sim \rho_g$ (in the mid-plane) may not robustly predict the onset of strong clumping of solids by the streaming instability and the ensuing planetesimal formation, where $\rho_p$ and $\rho_g$ are the local densities of the particles and the gas, respectively.
It appears that the criterion should instead depend on a combination of the solid abundance $Z$ and the radial diffusion of the particles driven by the turbulence.

Another important implication of the behavior of non-ideal protoplanetary disks is for pebble accretion onto growing planetesimals \citep{LJ12,JL17}.
In the regime of Bondi accretion (for smaller planetesimals), the accretion rate of pebbles inversely depends on the velocity dispersion of the pebbles.
We find in Section~\ref{SS:rcd} that the velocity dispersion of particles of $\tau_s = 0.1$ in MHD flow is comparable to or more than the difference between the gas and Keplerian velocities, which seems significant in this regime.
In the regime of Hill accretion (for larger planetesimals), the accretion rate of pebbles inversely depends on their scale height, when the scale height is greater than the Hill radius \citep{ML15,XBM17}.
In light of the appreciable difference in the mid-plane between the weak shear viscosity and the much stronger diffusion driven by velocity fluctuations in the dead zone of the protoplanetary disk, as found in this work, such a distinction should be considered in future works on pebble accretion in this regime in order to obtain a more realistic scale height of the pebbles along with viscous evolution of the disk.

Finally, we note that this work does not include additional non-ideal MHD effects, such as ambipolar diffusion or Hall drift, which can allow driving of a disk wind \citep{xB14,LKF14,GT15}.
Nevertheless, significant gas velocity dispersion near the mid-plane was still found in disk wind models including ambipolar diffusion \citep{SB13} as well as Hall drift \citep{xB15}, as compared to purely hydrodynamical streaming turbulence.
Moreover, \cite{ZSB15} found anomalous anisotropic diffusion in MHD turbulence controlled by ambipolar diffusion.
The result is a layer of particles significantly thicker than expected from the accretion shear stress \citep[see also][however]{RL18}.
\cite{XBM17} confirmed this result by showing significantly more depressed accretion stress than vertical diffusion of particles in ambipolar diffusion dominated flow, as compared to ideal-MHD models.
Therefore, the solid particle distribution appears to be regulated by anisotropic velocity fluctuations, whether the disk is controlled by Ohmic resistance or ambipolar diffusion.
It remains to be determined how effectively the streaming instability can concentrate solid particles in the latter case.

\acknowledgments

We thank Oliver Gressel, Michiel Lambrechts, Satoshi Okuzumi, and Zhaohuan Zhu for their discussion of this work.
We also thank the anonymous reviewer for their useful comments, which helped improve the accuracy of our manuscript.
We acknowledge PRACE for awarding us access to MareNostrum at Barcelona Supercomputing Center (BSC), Spain.
Part of the simulations and analyses were performed on resources provided by the Swedish National Infrastructure for Computing (SNIC) at PDC and LUNARC.
CCY and AJ acknowledge support from the European Research Council (ERC Starting Grant 278675-PEBBLE2PLANET and ERC Consolidator Grant 724687-PLANETESYS).
MMML was partly supported by NASA grant NNX14AJ56G.
AJ is grateful for additional support from the KAW Foundation (grant 2012.0150) and the Swedish Research Council (grant 2014-5775).

\software{The \pc\ (\url{http://pencil-code.nordita.org/})}

\appendix
\section{INTEGRATION OF STIFF OHMIC RESISTANCE} \label{S:isor}

With our choice of the magnetic Reynolds number $\ReM = 1$ in the mid-plane, the term for the Ohmic resistance in Equation~\eqref{E:mhd_ind} is particularly stiff.
The term dominates in the Courant condition and makes the explicit integration of our system impractical.
Therefore, we have devised a numerical algorithm, which is distinct from the ``super-time-stepping scheme'' \citep{AAG96} often adopted in the literature, to relieve the time-step constraint due to this term, as described below.

\subsection{The Algorithm} \label{SS:alg}

First, we rewrite the Ohmic resistance in the induction Equation~\eqref{E:mhd_ind} as follows:
\begin{align}
\mu_0\eta(z)\vec{J}
&= \eta(z)\curl{\curl{\vec{A}}}\nonumber\\
&= \eta(z)\left(\del\divgc{\vec{A}} - \laplacian{\vec{A}}\right)\nonumber\\
&= \del\left[\eta(z)\divgc{\vec{A}}\right]
    - (\divgc{\vec{A}})\del\eta(z)
    - \eta(z)\laplacian{\vec{A}},
\end{align}
The first term does not affect the evolution of the magnetic field $\vec{B}$ and can be removed by an appropriate gauge transformation.
The second and the third terms contain first and second derivatives of the magnetic vector potential $\vec{A}$, respectively.
The latter is the stiff term we aim to treat, and hence we operator split it out from Equation~\eqref{E:mhd_ind}, leading to the equations
\begin{align}
\pder{\vec{A}}{t}
    &= \frac{3}{2}\OmegaK x\pder{\vec{A}}{y}
    + \frac{3}{2}\OmegaK A_y\unitvec_x
    + \vec{u}\times\left(\vec{B} + \vec{B}_\mathrm{ext}\right)
    + (\divgc{\vec{A}})\del\eta(z),\label{E:os1}\\
\pder{\vec{A}}{t}
    &= \eta(z)\laplacian{\vec{A}}.\label{E:os2}
\end{align}
We integrate Equation~\eqref{E:os1} as usual in the \pc{} with finite differences and the Runge--Kutta method.
As long as the magnitude of $\del\eta$, which acts as an additional advection speed for $\vec{A}$, is not comparable with or significantly larger than the speed of sound, there exists no penalty in time steps in Equation~\eqref{E:os1} with explicit integration.
Finally, given that the resistivity $\eta(z)$ we use in our models only varies vertically, we can further dimensionally split Equation~\eqref{E:os2} into horizontal and vertical directions, resulting in the equations
\begin{align}
\pder{\vec{A}}{t}
    &= \eta(z)\left(\pdder{\vec{A}}{x} +
                    \pdder{\vec{A}}{y}\right),\label{E:os2h}\\
\pder{\vec{A}}{t}
    &= \eta(z)\pdder{\vec{A}}{z},\label{E:os2v}
\end{align}
respectively.

We integrate Equation~\eqref{E:os2h} as follows.
At any given vertical position $z$, Equation~\eqref{E:os2h} is a diffusion equation with a constant diffusion coefficient $\eta(z)$.
Therefore, it can be solved by the classic technique of Fourier transforms, under the assumption of periodic boundary conditions in both $x$ and $y$.
For the local-shearing-sheet approximation, we resort to the same technique already implemented in the \pc{} for the Poisson solutions of the gravitational potential \citep{JO07,YMM09,YMM12}.
This technique uses additional forward and inverse steps of Fourier interpolation to recover periodicity in the radial direction.
In any case, the Fourier solutions are analytical and hence the time step is not limited in this step.

Special care needs to be taken to integrate Equation~\eqref{E:os2v}.
It is a one-dimensional diffusion equation with spatially varying diffusion coefficient, and the technique of Fourier transforms does not apply in this case.
At any given horizontal position $(x, y)$, we adopt an implicit approach and discretize each component of Equation~\eqref{E:os2v} with second-order accuracy:
\begin{equation}
A_k^{(n+1)} =
A_k^{(n)} +
\frac{\eta(z_k)\Delta t}{2\Delta z^2}\left[
    \left(A_{k-1}^{(n+1)} - 2A_k^{(n+1)} + A_{k+1}^{(n+1)}\right) +
    \left(A_{k-1}^{(n)} - 2A_k^{(n)} + A_{k+1}^{(n)}\right)
\right],\label{E:os2vd}
\end{equation}
where $A_k^{(n)}$ is the specified component of the vector potential $\vec{A}$ at time $t^{(n)}$ and position $(x,y,z_k)$, $\Delta t \equiv t^{(n+1)} - t^{(n)}$ is the time step, and $\Delta z$ is the vertical cell size, assumed to be constant.
In combination with the vertical boundary conditions and one ghost cell on each side, Equation~\eqref{E:os2vd} constitutes a tridiagonal\footnote{For periodic boundary conditions in the vertical direction, the two off-diagonal corners of the coefficient matrix are also nonzero, which is known as a cyclic tridiagonal system.  A special numerical method to solve this system exists \citep[see, e.g.,][Section~2.7.2]{PT07}, which we use for the convergence study in the following section.} linear system of equations for $A_k^{(n+1)}$ and can be solved efficiently by the standard Gaussian elimination method.
With this implicit approach, the diffusion operator does not limit the time step either in this step.

\subsection{Damped Alfv\'{e}n Waves}

To validate the algorithm described in Section~\ref{SS:alg}, we resort to damped Alfv\'{e}n waves.
We adopt a cubic periodic Cartesian box of size $L$ with an incompressible fluid of density $\rho_0$.
The fluid has a constant kinematic viscosity of $\nu$ and a constant magnetic diffusivity $\eta$ with $\nu = \eta$, and hence the diffusion time is $\tau = L^2 / \nu = L^2 / \eta$.
It is threaded with an external uniform magnetic field of $\Bext = (B_0 / 3)(2\unitvec_x + 2\unitvec_y + \unitvec_z)$ such that the Alfv\'{e}n speed is $v_A = B_0 / \sqrt{\mu_0\rho_0} = 10^2 L/\tau$, where $\mu_0$ is the permeability.
Sinusoidal perturbations of wave vector $\vec{k} = (2\pi / L)(2\unitvec_x + 2\unitvec_y + \unitvec_z)$ that is parallel to $\Bext$ are initialized in the system.
The perturbation amplitude for the velocity is $\delta\vec{u} = 10^{-3}v_A\vec{w}$ and that for the magnetic field is $\delta\vec{B} = 10^{-3}B_0\vec{w}$, where $\vec{w} = \unitvec_x + \unitvec_y - 4\unitvec_z$, so that the energy equipartition and the solenoidal condition for both the velocity and the magnetic field are satisfied.
The solution for the evolution of the perturbations is analytically available \citep[Section~39]{sC61}, and we use it to measure the numerical errors involved in our algorithm. 

Because $\eta = \nu$, the stiffness of the resistive and viscous terms is the same.
They become stiff when the cell size $h \lesssim \nu / v_A = 10^{-2}L$.
Given that the viscous term has the same form as in Equation~\eqref{E:os2}, we use the same algorithm to integrate this term.

\begin{figure}
\plotone{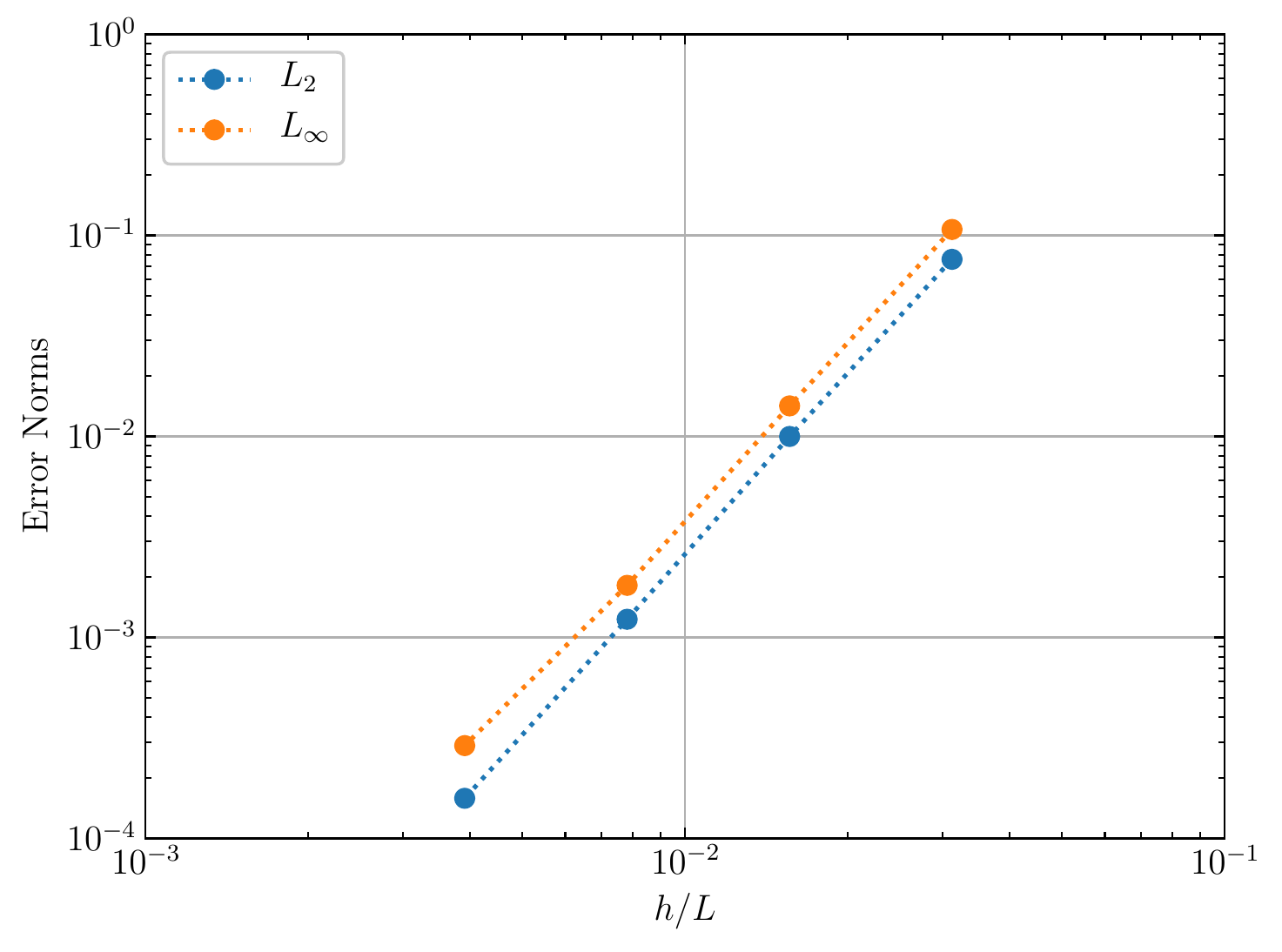}
\caption{The error norms in the $x$ component of the velocity as a function of cell size $h$ for the damped Alfv\'{e}n waves.
The errors are measured at $t = 0.01\tau$, where $\tau$ is the diffusion time, and they are normalized by the analytical amplitude at the time.
Third-order convergence is achieved.\label{F:convgc}}
\end{figure}

Figure~\ref{F:convgc} illustrates the convergence in the $x$ component of the velocity for this system using our algorithm.
We evolve the system for $0.01\tau$, and measure the resulting $L_2$ and $L_\infty$ norms against the analytical solution.
Both norms demonstrate a third-order convergence over the cell sizes from L / 32 down to L / 256, which covers the transition point where the resistive and viscous terms become stiff.


\end{document}